# ActorScript™ extension of C#®, Java®, Objective C®, JavaScript®, and SystemVerilog using iAdaptive™ concurrency for antiCloud™ privacy and security

**Carl Hewitt**

***This article is dedicated to Ole-Johan Dahl and Kristen Nygaard.***

**Message passing using types is the foundation of system communication:**

- Messages are the unit of communication
- Types enable secure communication Actors

ActorScript™ is a general purpose programming language for implementing iAdaptive™ concurrency that manages resources and demand. It is differentiated from previous languages by the following:

- Universality
    - Ability to directly specify exactly what Actors can and cannot do
    - Everything is accomplished with message passing using types including the very definition of ActorScript itself.
        - Messages can be directly communicated without requiring indirection through brokers, channels, class hierarchies, mailboxes, pipes, ports, queues *etc.* Programs do not expose low-level implementation mechanisms such as threads, tasks, locks, cores, *etc*. Application binary interfaces are afforded so that no program symbol need be looked up at runtime. Functional, Imperative, Logic, and Concurrent programs are integrated.
        - A type in ActorScript is an interface that does not name its implementations (contra to object-oriented programming languages beginning with Simula that name implementations called "classes" that are types). ActorScript can send a message to any Actor for which it has an (imported) type.

- Concurrency can be dynamically adapted to resources available and current load.
- Safety, security and readability
  - Programs are *extension invariant*, *i.e.*, extending a program does not change the meaning of the program that is extended.
  - Applications cannot directly harm each other.
  - Variable races are eliminated while allowing flexible concurrency.
  - Lexical singleness of purpose. Each syntactic token is used for exactly one purpose.
- Performance[i]
  - Imposes no overhead on implementation of Actor systems in the sense that ActorScript programs are as efficient as the same implementation in machine code. For example, message passing has essentially same overhead as procedure calls and looping.
  - Execution dynamically adjusted for system load and capacity (*e.g.* cores)
  - Locality because execution is not bound by a sequential global memory model
  - Inherent concurrency because execution is not limited by being restricted to communicating *sequential* processes
  - Minimize latency along critical paths

ActorScript attempts to achieve the highest level of performance, scalability, and expressibility with a minimum of primitives.

**C#** is a registered trademark of Microsoft, Inc.
**Java** and **JavaScript** are registered trademarks of Oracle, Inc.
**Objective C** is a registered trademark of Apple, Inc.
***Computer software should not only work; it should also appear to work.***[1]

**Introduction**

ActorScript is based on the Actor mathematical model of computation that treats "*Actors*" as the universal primitives of concurrent digital computation

---

[i] Performance can be tricky as illustrated by the following:

- "Those who would forever give up correctness for a little temporary performance deserve neither correctness nor performance." [Philips 2013]
- "The key to performance is elegance, not battalions of special cases" [John Bentley]
- "If you want to achieve performance, start with comprehensible." [Philips 2013]
- Those who would forever give up performance for a feature that slows everything down deserve neither the feature nor performance.

[Hewitt, Bishop, and Steiger 1973; Hewitt 1977; Hewitt 2010a]. Actors have been used as a framework for a theoretical understanding of concurrency, and as the theoretical basis for several practical implementations of concurrent systems.

### ActorScript

ActorScript is a general purpose programming language for implementing massive local and nonlocal concurrency.

This paper makes use of the following typographical conventions that arise from underlying namespaces for types, messages, language constructs, syntax categories, *etc.*[i]

- type identifiers (*e.g.,* Integer)
- program variables (*e.g.,* aBalance)
- message names (*e.g.,* getBalance)
- reserved words[2] for language constructs (*e.g.,* **Actor**)
- structures (*e.g.,* [ and ])
- argument keyword (e.g., to )
- logical variables (e.g., *x*)
- comments in programs (e.g. /* this is a comment */ )

There is a diagram of the syntax categories of ActorScript in an appendix of this paper in addition to an appendix with an index of symbols and names along with an explanation of the notation used to express the syntax of ActorScript.[3]

### Actors

ActorScript is based on the Actor Model of Computation [Hewitt, Bishop, and Steiger 1973; Hewitt 2010a] in which all computational entities are Actors and all interaction is accomplished using message passing.

The Actor model is a mathematical theory that treats "*Actors*" as the universal primitives of digital computation. The model has been used both as a framework for a theoretical understanding of concurrency, and as the theoretical basis for several practical implementations of concurrent systems. Unlike previous models of computation, the Actor model was inspired by physical laws. The advent of massive concurrency through client-cloud computing and many-core computer architectures has galvanized interest in the Actor model.

---

[i] The choice of typography in terms of font and color has no semantic significance. The typography in this paper was chosen for pedagogical motivations and is in no way fundamental. Also, only the abstract syntax of ActorScript is fundamental as opposed to the surface syntax with its many symbols, e.g., ↦, etc.

An Actor is a computational entity that, in response to a message it receives, can concurrently:

- send messages to addresses of Actors that it has
- create new Actors
- for a serialized Actor, designate how to handle the next message it receives.

There is no assumed order to the above actions and they could be carried out concurrently. In addition two messages sent concurrently can be received in either order. Decoupling the sender from communication it sends was a fundamental advance of the Actor model enabling asynchronous communication and control structures as patterns of passing messages.

The Actor model can be used as a framework for modeling, understanding, and reasoning about, a wide range of concurrent systems. For example:

- Electronic mail (e-mail) can be modeled as an Actor system. Mail accounts are modeled as Actors and email addresses as Actor addresses.
- Web Services can be modeled with endpoints modeled as Actor addresses.
- Object-oriented programing objects with locks (e.g. as in Java and C#) can be modeled as Actors.

Actor technology will see significant application for integrating all kinds of digital information for individuals, groups, and organizations so their information usefully links together. Information integration needs to make use of the following information system principles:

- ***Persistence***. *Information is collected and indexed.*
- ***Concurrency***: *Work proceeds interactively and concurrently, overlapping in time.*
- ***Quasi-commutativity***: *Information can be used regardless of whether it initiates new work or becomes relevant to ongoing work.*
- ***Sponsorship***: *Sponsors provide resources for computation, i.e., processing, storage, and communications.*
- ***Pluralism***: *Information is heterogeneous, overlapping and often inconsistent. There is no central arbiter of truth.*
- ***Provenance***: *The provenance of information is carefully tracked and recorded.*

The Actor Model is designed to provide a foundation for inconsistency robust information integration.

**Syntax**

To ease interoperability, ActorScript uses an intersection of the orthographic conventions of Java, JavaScript, and C++ for words[i] and numbers.

**Expressions**

ActorScript makes use of a great many symbols to improve readability and remove ambiguity. For example the symbol "▮" is used as the top level terminator to designate the end of input in a read-eval-print loop. An Integrated Development Environment (IDE) can provide a table of these symbols for ease of input as explained below:[ii]

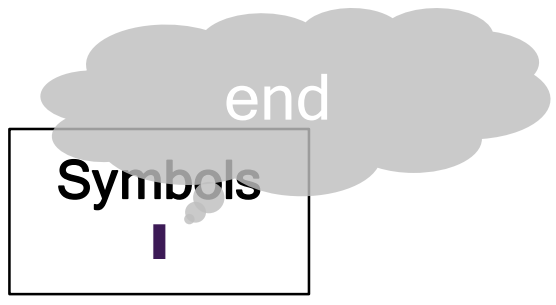

Expressions evaluate to Actors. For example, 1+3▮[iii] is equivalent[iv] to 4▮.

Parentheses "(" and ")" can be used for precedence. For example using the usual precedence for operators, 3*(4+2)▮ is equivalent to 18▮, while 3*4+2▮ is equivalent to 14▮,

Identifiers, e.g., x, are expressions that can be used in other expressions. For example if x is 1 then x+3▮ is equivalent to 4▮. The formal syntax of identifiers is in the following end note: **4.**

**Types**

Types are Actors. In this paper, Types are shown in green, *e.g.,* **Integer**.

The formal syntax for types is in the following end note: **5.**

**Definitions,** *i.e.,* ≡

A simple definition has the name to be defined followed by "≡" followed by the definition. For example, x:**Integer**≡3▮ defines the identifier x to be of type **Integer** with value 3.

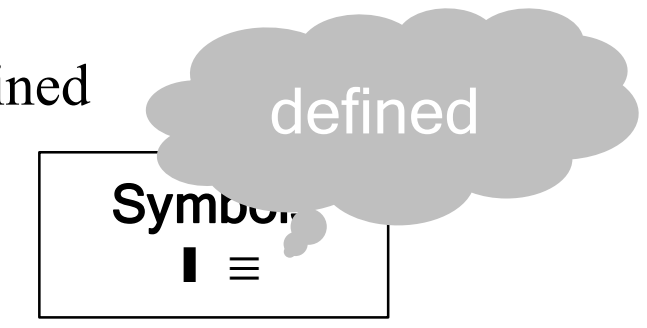

The formal syntax of a definition is in the end note: **6.**

---

[i] sometimes called "names"

[ii] Furthermore, all special symbols have ASCII equivalents for input with a keyboard. An IDE can convert ASCII for a symbol equivalent into the symbol. See table in an appendix to this article.

[iii] An IDE can provide a box with symbols for easy input in program development. The grey callout bubble is a hover tip that appears when the cursor hovers above a symbol to explain its use.

[iv] in the sense of having the same value and the same effects

**Interfaces for procedures,** *i.e.,* **Interface { [ ]↦ }**

A procedure interface is used to specify the types of messages that a procedure Actor can receive. The syntax is "**Interface**" followed by an interface identifier, and procedure signatures in parentheses separated by commas. A procedure signature consists of a message signature with argument types delimited by "[" and "]", followed by "↦", and a return type.[i] An alternative syntax (which is more like Java) is that a procedure signature can be written as a return type followed by "↤", and message signature with argument types delimited by "[" and "]".

For example, the interface[ii] for the overloaded[7] procedure type IntegerToIntegerAndVectorToVector that takes an Integer argument to return an Integer value and a Vector argument and to return a Vector can be constructed as follows:[8]

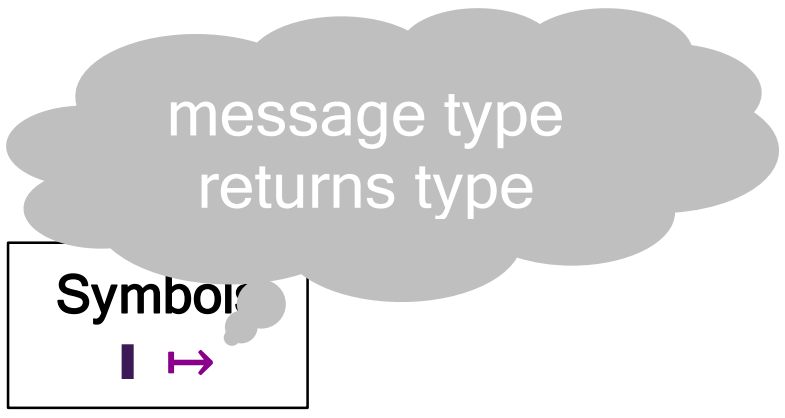

**Interface** IntegerToIntegerAndVectorToVector
{[Integer]↦ Integer, // equivalently Integer↤[Integer]
[Vector]↦ Vector}❚

For security reasons, the type IntegerToIntegerAndVectorToVector is *different* from the type constructed below:[iii]

**Interface** VectorToVectorAndIntegerToInteger
{[Vector]↦ Vector,
[Integer]↦ Integer}❚

The formal syntax of a procedure interface is in the following end note: **9.**

**Procedures,** *i.e.,* **Actor implements [ ]→ , ¶ and §**

A procedure has message formal parameters delimited by "[" and "]" followed by "→" and then the expression to be computed.[iv] For example,

---

[i] Since communicating using messages is crucial for Actor systems, messages are shown in red in this article. The choice of color has no semantic significance.

[ii] Every interface is a type.

[iii] Merely, having procedures with the same signatures does not make IntegerToIntegerAndVectorToVector the same type as VectorToVectorAndIntegerToInteger.

[iv] Note the following crucial differences (recalling that font, color, and capitalization are of no semantic significance for identifiers although words with different capitalization are different identifiers):

[n:**Integer**]→n+n▮ is a (unnamed) procedure that given a message with an integer number, n, returns the number plus itself.

Procedures can be overloaded using "**Actor implements**", followed by a type, followed by "**using**", followed by a list of procedures separated by "¶" and terminated by "§".[i] For example, in the following Double is defined to implement **IntegerToIntegerAndVectorToVector**.

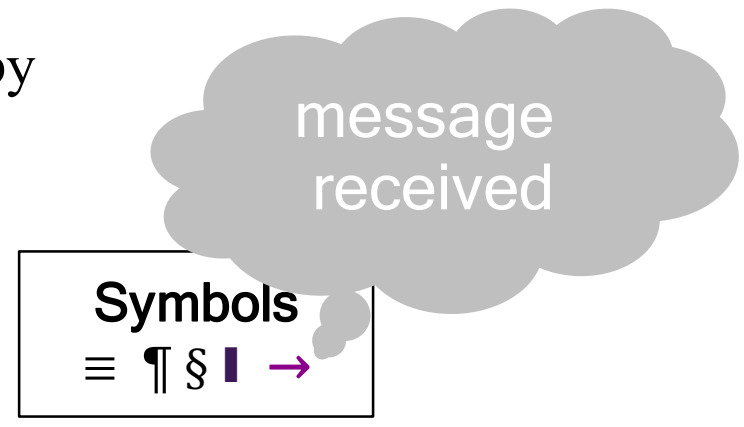

```
Double ≡ Actor
          implements IntegerToIntegerAndVectorToVector using
            [n:Integer]→ n+n¶          // integer addition
            [v:Vector]→ v+v §▮         // vector addition
```

The formal syntax of procedures is in the end note: **10.**

**Sending messages to procedures,** *i.e.*, ▪[ ]

Sending a message to a procedure (*i.e.* "calling" a procedure with arguments) is expressed by an expression that evaluates to a procedure followed by "▪"[11] followed by a message with parameter expressions delimited by "[" and "]". For example, Square▪[2+1]▮ means send Square[ii] the message [3]. Thus Square▪[2+1]▮ is equivalent to 9▮.

The formal syntactic definition of procedural message sending is in the end note: **12.**

---

- [Integer]↦Integer is a procedure signature type and *not* a procedure. It is a procedure type for a procedure that takes an Integer argument and returns an Integer.
- [Integer]→Integer is a procedure and *not* a type. It is the "identity" procedure of one argument that always returns the argument.

[i] Since both procedures and implementations can be quite large, an IDE can use these special symbols to provide additional help.

[ii] As a convenience, the procedure Square can be defined to implement the type [Integer]↦Integer as follows: Square▪[x:Integer]:Integer ≡ x*x▮

**Patterns**

Patterns are fundamental to ActorScript. For example,

- 3 is a pattern that matches 3
- "abc" is a pattern that matches "abc".
- _ is a pattern that matches anything[i]
- _:**Integer** is a pattern that matches any **Integer**
- $$x is a pattern that matches the value of x.
- $$(x+2) is a pattern that matches the value of the expression x+2.
- < 5 is a pattern that matches an integer less than 5
- **x suchThat** Factorial.**[x]>120** is a pattern that matches an integer whose factorial is greater than 120

Identifiers[ii] can be bound using patterns as in the following examples:

- x is a pattern that matches "abc" and binds x to "abc"
- x:**Integer** is a pattern that does not match "abc" because "abc" is not an integer
- x:**Integer** is a pattern that matches 3 and binds x to 3

**Cases,** *i.e.*, **� ⦂ , ⦂ ⍰**

Cases are used to perform conditional testing. In a Cases Expression, an expression for the value on which to perform case analysis is specified first followed by "**�**"[iii] and then followed by a number of cases such that each case is separated from the next by "**,**" and cases are terminated by "⍰".[13] A case consists of

- a pattern followed by "⦂" and an expression to compute the value for the case. *All of the patterns before an* **else** case must be disjoint; *i.e.*, it must not be possible for more than one to match.
- optionally (at the end of the cases) *one or more* of the following cases: "**else**" followed by an optional pattern, "⦂", and an expression to compute the value for the case. An **else** case applies *only* if none of the patterns in the preceding cases[iv] match the value on which to perform case analysis.

---

[i] e.g., _ matches 7

[ii] An identifier is a name that is used in a program to designate an Actor

[iii] "**�**" is fancy typography for "**?**"

[iv] *including* patterns in previous else cases

As an arbitrary example purely to illustrate the above, suppose that the procedure Random is of type **[ ]↦Integer** in the following example:

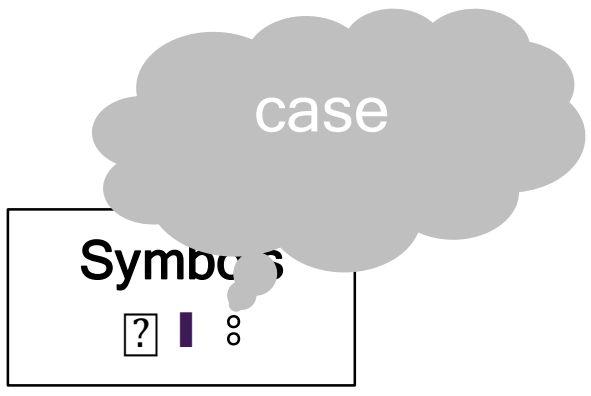

Random.[ ] �
  0 ⦂     // Random.[ ] returned 0[i]
   **Throw**[ii] RandomNumberException[ ],
        // throw an exception because Fibonacci.[0] is undefined
  1 ⦂                // Random.[ ] returned 1
   **6,**              // the value of the cases expression is 6
  **else** y **thatIs** < 5 ⦂
        // Random.[ ] returned y that is not 0 or 1 and is less than 5
   Fibonacci.[y], // return Fibonacci of the value returned by Random.[ ]
  **else** z ⦂  // Random.[ ] returned **z** that is not 0 or 1 and is not less than 5
   Factorial.[z] ⍰▮ // return Factorial of the value returned by Random.[ ]

The formal syntax of cases is in the following end note: **14.**

**Binding locals, *i.e.*, Let {← }**

Local identifiers can be bound using "**Let**" followed by a pattern, "←", an expression for the Actor to be matched, "**,**", and an expression in which the identifiers can be used to compute an Actor. For example, [“G”, “F”, “F”]▮ could be written as follows:

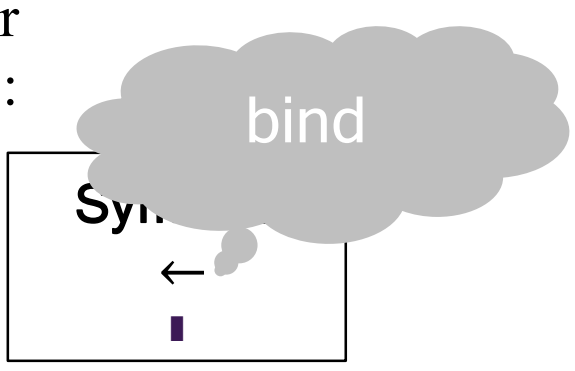

  **Let** x ← “F”,        // x is “F”
   [“G”, x, x]▮

---

[i] As is standard, ActorScript uses the token “//” to begin a one-line comment. In this article, comments are depicted in gray font.
[ii] Reserved words are shown in bold black.

Dependent bindings (in which each can depend on previous ones) can be accomplished using "[" followed by bindings separated using "," terminated by "]". Also, a binding can accomplished using a list pattern. For example, ["L", ["H", "F"], ["K", "F"]]▮ could be written as follows:

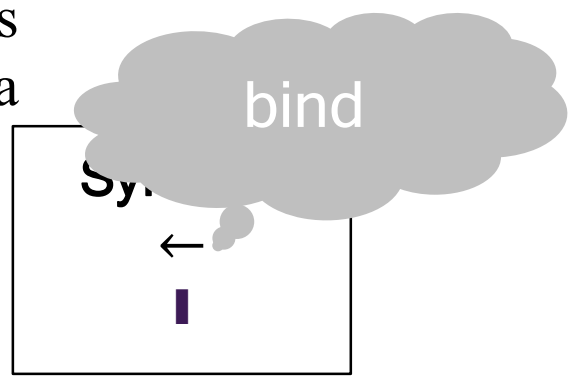

```
Let [x ←"F",                          // x is "F"
     y ←["G", x, x],        // y is ["G", "F", "F"]
     [u, v] ←[["H", x], ["K", x]]]   // u is ["H", "F"] and v is ["K", "F"]
  ["L", u, v]▮
```

Also, multiple results can be bound. For example

```
Let [quotient, remainder] ← QuotientRemainder 7/3,
                                        // quotient is 2 and remainder is 1
  quotient+remainder▮                   // returns 3
```

The formal syntax of bindings is in the following end note: **15.**

**Actor expressions with Assignable Variables, *i.e.,* Actor and ≔**

Using the expressions introduced so far, actors do not change. Mutable Actors are introduced below.

An Actor can be created using "**Actor**" optionally followed by the following:

- constructor name with formal arguments delimited using brackets
- declarations of variables[i]
- implementations of interface(s).

Reserved words (*e.g.*, **Actor**) are case sensitive. Furthermore, an infix reserved word is always lower case.

Message handlers in an Actor execute mutually exclusively. In this paper assignable variables are colored orange, which by itself has no semantic significance, i.e., printing this article in black and white does not change any meaning. The use of assignments is strictly controlled in order to achieve better structured programs.[16]

---

[i] each variable declaration followed by a comma

Below is an example of an account which provides the ability to get the current balance, deposit an amount, and withdraw an amount:[17]

Symbols
≡ → ⦂ ⍰ ¶ § ▮ ≔

assignment

```
Actor SimpleAccount[startingBalance:Euro]
   myBalance ≔ startingBalance,
    // myBalance is an assignable variable initialized with startingBalance
   implements Account[i] using
     getBalance[ ] →  myBalance¶
     deposit[anAmount:Euro] →
        Void                              // return Void
           afterward  myBalance ≔ myBalance+anAmount¶
                        // the next  message is processed with
                               //  myBalance reflecting the deposit
     withdraw[anAmount:Euro] →
        (amount > myBalance) �
          True ⦂ Throw  OverdrawnException[ ],
          False ⦂ Void                  // return Void
                   afterward myBalance ≔ myBalance−anAmount ⍰§▮
                     // the next message is processed with myBalance
```

The formal syntax of **Actor** expressions is in the following end note: **18.**

A message handler signature consists of a message name followed by argument types delimited by "[" and "]", "↦", and a return type. An alternative syntax (which is more like Java) is that a message handler signature can be written as a return type followed by "↤", message name, and argument types delimited by "[" and "]".

The formal syntactic definition of named-message sending is in the following end note: **19**

### Continuations

Continuations are used in ActorScript to linearize computation and to increase referential transparency of variables.[20] Regions highlighted in yellow above are continuations.

---

[i] **Interface** Account{getBalance[ ]↦Euro,          // equivalently Euro↤getBalance[ ]
deposit[Euro]↦Void,
withdraw[Euro]↦Void]}▮

For example, the following sub-continuation in the **withdraw** handler

**Void afterward** **myBalance** ≔ **myBalance**+anAmount returns **Void** and updates **myBalance** for the *next* message received.

By linearizing computation, a continuation prevents default concurrency and consequently variable data races are impossible.

The formal syntax of continuations is in the following end note: **21.**

**Antecedents, Preparations, and Concurrency, *i.e.*, ; and ⦷**

An expression can be annotated for concurrent execution by preceding it with "⦷" indicating that the following expression should be considered for concurrent execution if resources are available. For example ⦷Factorial▪[1000]+⦷Fibonacci▪[2000]▮ is annotated for concurrent execution of Factorial▪[1000] and Fibonacci▪[2000] both of which *must* complete execution. This does not require that the executions of Factorial▪[1000] and Fibonacci▪[2000] actually overlap in time.

The formal syntax of explicit concurrency is in the following end note: **22.**

Concurrency can be controlled using preparation that is expressed in a continuation using "**Do**" followed by a preparatory expressions, "⬣" and an expression that proceeds only *after* the preparations have been completed.

The following expression creates an account anAccount with initial balance €5 and then concurrently withdraws €1 and €2 in preparation for reading the balance:

Euro Symbol ← €

**Let** anAccount ← SimpleAccount▪[€6], //€ is a reserved prefix operator
**Do** {⦷anAccount▪**withdraw**[€1],
⦷anAccount▪**withdraw**[€2]}⬣
// proceed only after both of the
// withdrawals have been acknowledged
anAccount▪**getBalance**[ ]▮

The above expression returns €3.

Operations are quasi-commutative to the extent that it doesn't matter in which order they occur. Quasi-commutativity can be used to tame indeterminacy.

The formal syntax of compound expressions is in the following end note: **23**

**Swiss cheese**

Swiss cheese [Hewitt and Atkinson 1977, 1979; Atkinson 1980][24] is a generalization of mutual exclusion with the following goals:

- *Generality:* Ability to conveniently program any scheduling policy
- *Performance:* Support maximum performance in implementation, e.g., the ability to minimize locking and to avoid repeatedly recalculating a condition for proceeding.
- *Understandability:* Invariants for the variables of a mutable Actor should hold whenever entering or leaving the cheese.
- *Modularity:* Resources requiring scheduling should be encapsulated so that it is impossible to use them incorrectly.

Message handlers in an Actor execute mutually exclusively while in the "cheese", *i.e.*, at most one activity can execute in the cheese at a time. However, there can be "holes" in the cheese to permit other activities to happen and then continue execution. This is achieved using "string bean style"[25] to control visibility of effects (*e.g.* assignments) and to enforce sequencing moving in and out of the cheese.

In the examples below, holes in the cheese are highlighted in grey and queues are shown in orange. The color has no semantic significance. In addition to reserved words being case sensitive, if a reserved word begins a phrase (*e.g.* **Enqueue**) then it always capitalized and infix reserved words within the phrase (*e.g.* **for, permit, afterward,** etc.) are lower case.

**A variable can change only as follows:**[i]

- just after leaving the cheese or after an internal delegated operation.
- when cheese is (re-)entered, a variable has the value when the cheese was last left.

---

[i] Consequently, variable races are *impossible*.

Below is an implementation of a **Gate** that suspends those who send a passThru[ ] and whenever the gate receives an open[ ] message, then those already waiting are allowed to pass through.

| Symbols |
|---|
| ≡ → |
| ⍰ ¶ § ▮ |

**Actor** PassThruWhenOpenedGate[ ]<br>
  **queue** aQueue<br>
    // declare aQueue to be a queue for activities, which is initially empty<br>
  **implements** Gate **using**<br>
    passThru[ ]→<br>
      Enqueue aQueue⬣<br>
          // Enqueue this activity in aQueue and then leave cheese<br>
        Void       // when resumed return Void and<br>
            permit aQueue¶   // permit the first of aQueue[26]<br>
    open[ ]→ Void     // return Void<br>
          permit aQueue§▮    // resume the first of aQueue[27]

The formal syntax of the above is in the following end note: **28**

The following is a state diagram of the above implementation PassThruWhenOpenedGate:

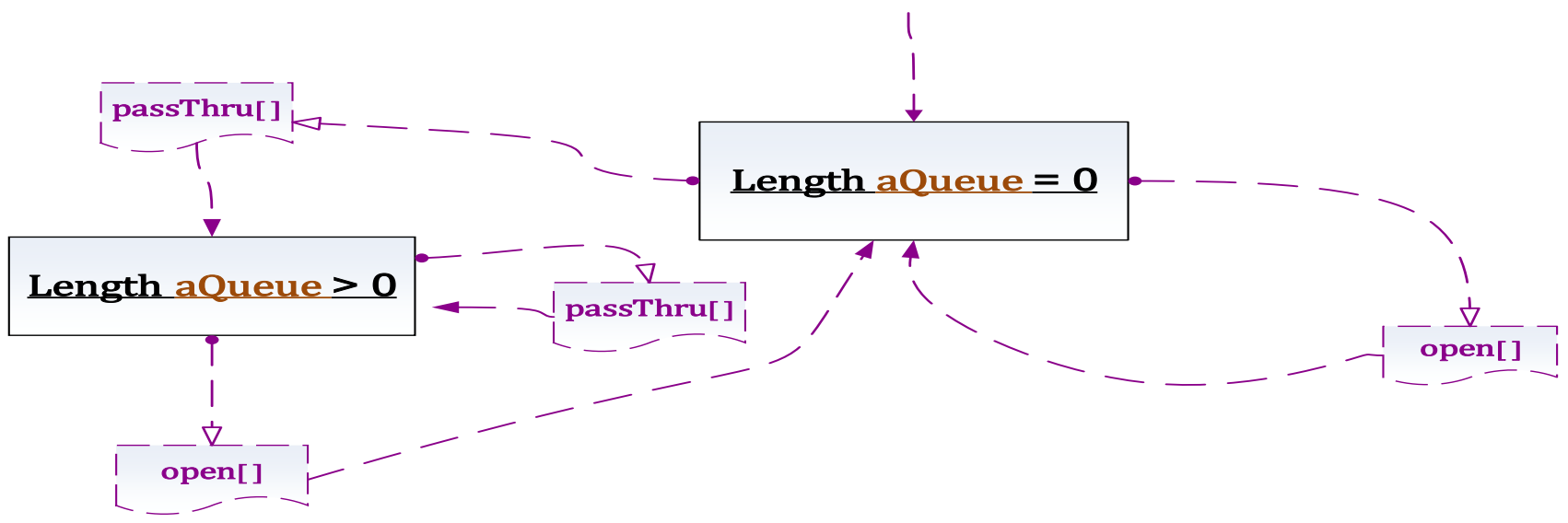

Below is an implementation of a **Gate**[i] that suspends those who send a passThrough[ ] until the gate is opened by receiving an open[ ] message and thereafter remains open:

| Symbols |
|---|
| ≡ → ⸰ |
| ⍰ ¶ § ▮ |

```
Actor OnetimeGate[ ]
    queue aQueue,
    opened ≔ False,
                    // opened is a local assignable variable that is initially False
    implements Gate using
        passThru[ ]→
            Do opened � False ⸰ Enqueue aQueue ⬣ Void ,
                            // if opened is False, then join aQueue and leave cheese
                             True ⸰ Void ⍰ ⬣
                              // if opened is True
                               // then proceed immediately without leaving cheese
              Precondition[29] opened,
                                    // opened must be True or an exception is thrown[30]
                 Void                           // return Void and
                    permit aQueue¶              // resume the first of aQueue[31]
        open[ ]→ Void                            // return Void and
                      permit aQueue // resume the first of aQueue[32] also
                           always opened ≔ True §▮
                                                // opened always assigned True
```

The following is a state diagram of the above implementation **OnetimeGate**:

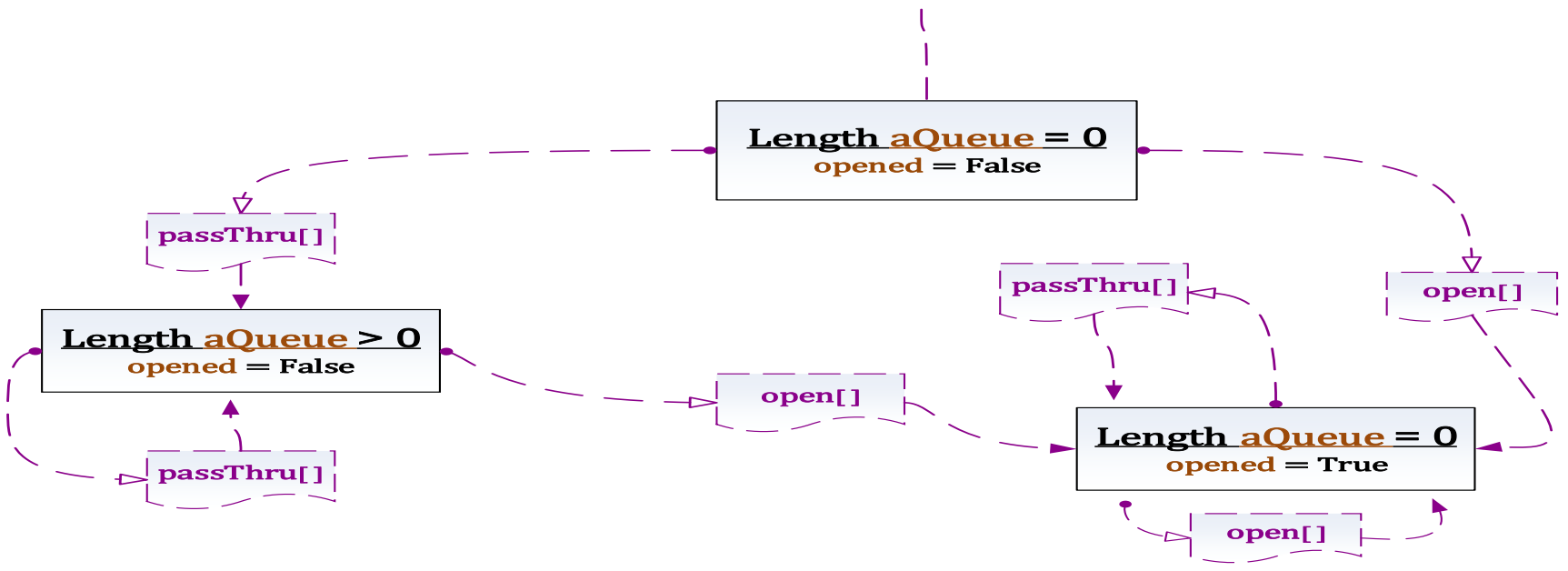

The formal syntax of the above is in the following end note: **33**

[i] Interface Gate {passThru[ ]↦ Void,
open[ ]↦ Void}▮

By contrast with the nondeterministic lambda calculus, there is an always-halting Actor that when sent a **start[ ]** message can compute an integer of unbounded size. This is accomplished by sending a counter that it creates both a **stop[ ]** message and a **go[ ]** message. The counter is created with an integer variable **currentCount** initially **0** and a Boolean variable **continue** that is initially **True** with the following behavior:

- When a **stop[ ]** message is received, set **continue** to **False** and return **currentCount**.
- When a **go[ ]** message is received:
  1. if **continue** is **True**, increment **currentCount** by 1 and send the counter a **go[ ]** message.
  2. if **continue** is **False**, return **Void**

By the Actor Model of Computation [Clinger 1981, Hewitt 2006], the above Actor will eventually receive the **stop[ ]** message and return an unbounded number.

| Symbols |
| --- |
| ≡ → ⦂ ← Ⓓ |
| ⍰ ¶ § ▮ |

As a convenience, a message can be delegated to this Actor by prefacing the message with "..".

```
Unbounded ≡
 start[ ]→   // a start message is implemented by
  Let aCounter ← SimpleCounter.[ ]   // let aCounter be a new Counter
   Do ⒹaCounter.go[ ],
                    // send aCounter a go message and concurrently
    ⒹaCounter.stop[ ]▮
                 // return the result of sending aCounter a stop message

Actor SimpleCounter[ ]
  count ≔ 0,                    // the variable count is initially 0
  continue ≔ True,
  implements Counter[34] using
    stop[ ] →
      count                         // return count
        afterward continue ≔ False¶
          // continue is updated to False for the next message received
    go[ ] →
      continue �
        True ⦂ Hole ..go[ ]           // send go[ ] to this counter after
             after count ≔ count+1 ,        // incrementing count
        False ⦂ Void ⍰§▮           // if continue is False, return Void
```

The formal syntax of the above is in the following end note: **35**

A barbershop[36] with two barbers can be implemented as follows:

```
Actor SimpleShop[waitingRoomCapacity:Integer,
                 firstBarber:Barber,
                 secondBarber:Barber]                  Symbols
                                                       ≡ → ⨾ ∧ ∨ ¬
                                                       ⍰ ¶ § ▮
  queue aQueue,
  firstBarberIsShaving ≔ False,      // initially neither barber is shaving
  secondBarberIsShaving ≔ False,
  implements BarberShop37 using
   visit[aClient:Client] →
     Do (Length aQueue) > waitingRoomCapacity �
          True ⨾ Throw WaitingRoomFull[ ],       // waiting room is full
          False ⨾ (firstBarberIsShaving ∧ secondBarberIsShaving) �
                  True ⨾ Enqueue aQueue⬣ Void,
                   // if both barbers are shaving then enqueue in aQueue
                   False ⨾ Void ⍰⍰⬣
       Precondition ¬firstBarberIsShaving ∨ ¬secondBarberIsShaving,
             // one of the barbers must be free or an exception is thrown38
         ¬firstBarberIsShaving �
            True ⨾                  // first barber is always preferred
              Hole firstBarber.shave[aClient]
                     // leave cheese while shaving
                                 // after recording that first barber is shaving
                 after firstBarberIsShaving ≔ True

                 returned� aTip ⨾
                    Do firstBarber.giveTip[aTip],
                      Void permit aQueue 39
                                always firstBarberIsShaving ≔ False⍰

                 threw permit aQueue
                          always firstBarberIsShaving ≔ False,
            False ⨾
              Hole secondBarber.shave[aClient]40
                 after secondBarberIsShaving ≔ True
                 returned�  aTip ⨾
                    Do secondBarber.giveTip[aTip],
                      Void permit aQueue 41
                          always
                            secondBarberIsShaving ≔ False ⍰

                 threw permit aQueue
                        always secondBarberIsShaving ≔ False ⍰§▮
```

The following is a state diagram of the above implementation Shop:

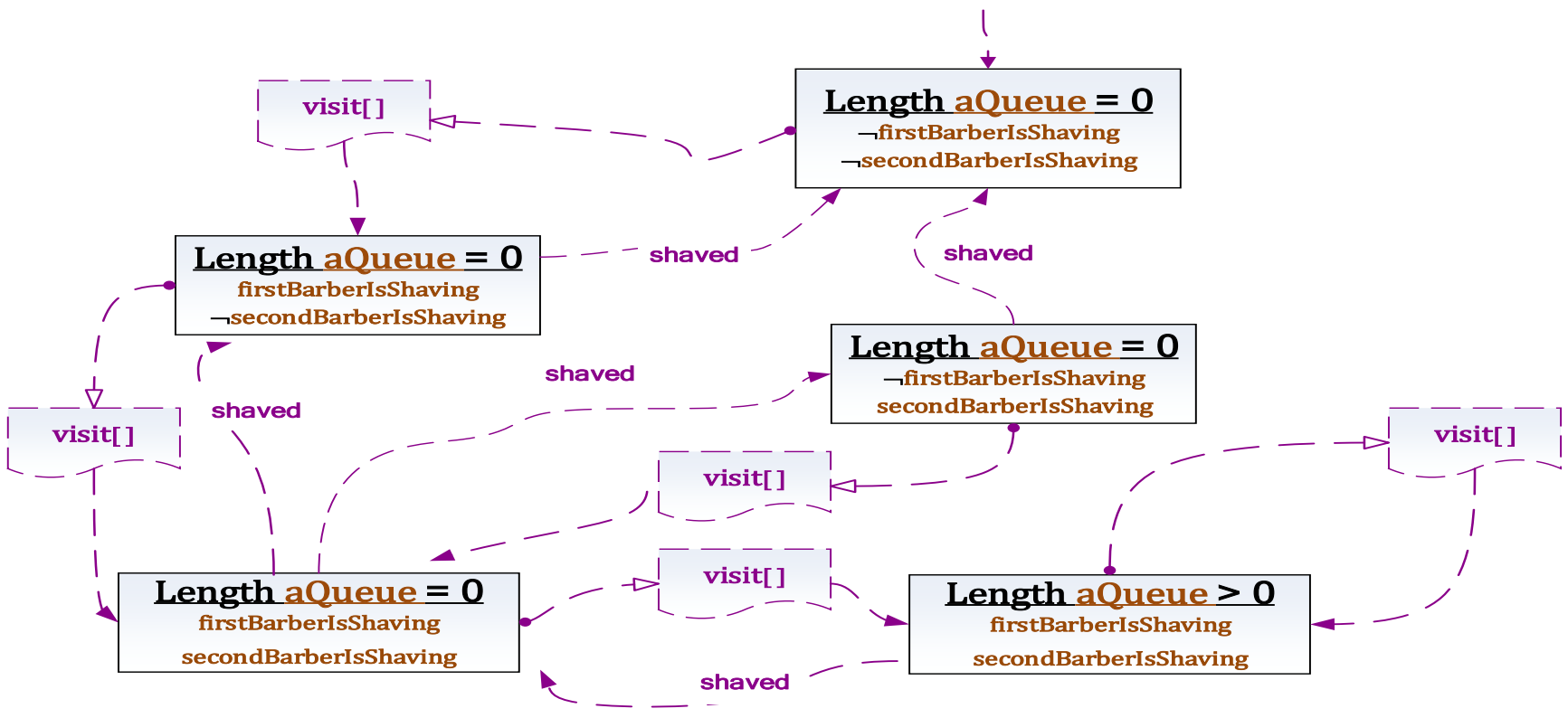

Concurrency control for readers and writers in a shared resource is a classic problem. The fundamental constraint is that multiple writers are not allowed to operate concurrently and a writer is not allowed operate concurrently with a reader.

**Swiss cheese with holes**

Below are two implementations of readers/writer guardians for a shared resource that implement different policies:[42]

1. *ReadingPriority:* The policy is to permit maximum concurrency among readers without starving writers.[43]
    a. When no writer is waiting, all readers start as they are received.
    b. When a writer has been received, no more readers can start.
    c. When a writer completes, all waiting readers start even if there are writers waiting.
2. *WritingPriority:* The policy is that readers get the most recent information available without starving writers.[44]
    a. When no writer is waiting, all readers start as they are received.
    b. When a writer has been received, no more readers can start.
    c. When a writer completes, just one waiting reader is permitted to complete if there are waiting writers.

The interface for the readers/writer guardian is the same as the interface for the shared resource:

**Interface** ReadersWriter {read[Query]↦ QueryResult,
write[Update]↦ Void}▮

State diagram of ReadersWriter implementations:

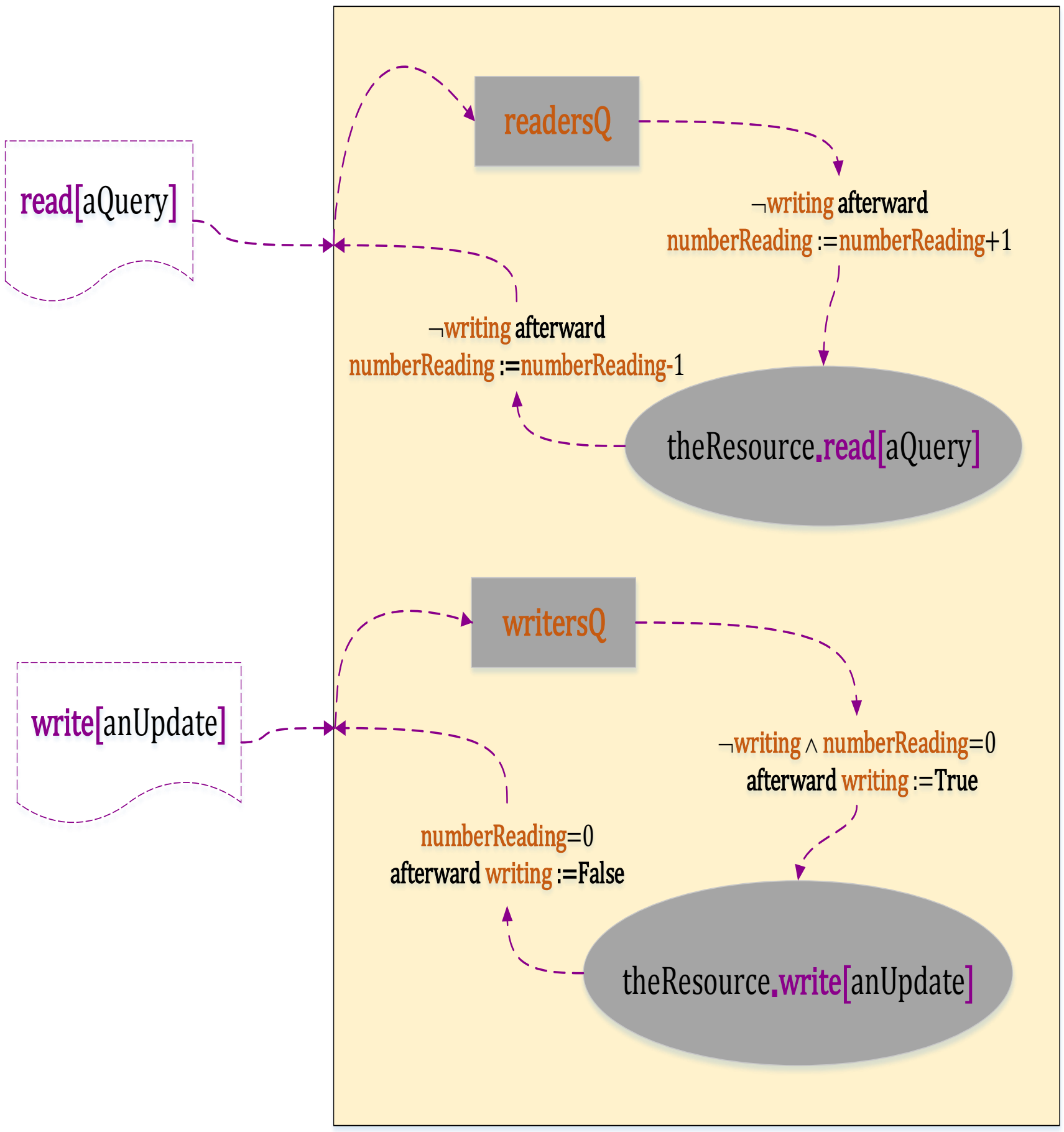

Note:

1. At most one activity is allowed to execute in the cheese.[i]
2. The cheese has holes.[ii]
3. The value of a variable[iii] changes only when leaving the cheese or after an internal delegated operation.[iv]

---

[i] Cheese is yellow in the diagram
[ii] A hole is grey in the diagram
[iii] A variable is orange in the diagram
[iv] Of course, other external Actors can change.

```
Actor ReadingPriority[theResource:ReadersWriter]
  queues {readersQ, writersQ}     // readersQ and writersQ are initially empty
  writing ≔ False,
  numberReading:(Integer thatIs ≧0) ≔ 0,
  implements ReadersWriter using
   read[query]→
     Do (writing ∨ ¬IsEmpty writersQ) �
          True ⦂ Enqueue readersQ⬢ Void   // leave cheese while in readersQ
                  backout (¬writing ∧ numberReading=0 ∧ IsEmpty readersQ) �
                               True ⦂ Void permit writersQ,
                               False ⦂ Void ⍰,
          False ⦂ Void ⍰⬢
         Precondition ¬writing,[45]
           Hole theResource.read[query]    // leave cheese while
                                   // reading after recording that another reader is reading
                 after permit readersQ always numberReading++  [46]
                 afterward
                     (IsEmpty writersQ) �
                        True ⦂ permit readersQ always numberReading––, [47]
                        False ⦂ numberReading �
                                  1 ⦂ permit writersQ always numberReading––,
                                  else ⦂ also numberReading–– ⍰ ⍰
   write[update]→
     Do numberReading>0∨¬IsEmpty readersQ ∨ writing ∨ ¬IsEmpty writersQ �
         True ⦂ Enqueue writersQ⬢ Void     // leave cheese while in writersQ
                      backout (IsEmpty writersQ ∧¬writing) �
                                True ⦂ Void permit readersQ,
                                 False ⦂ Void ⍰,
         False ⦂ Void ⍰⬢
        Precondition[48] numberReading=0 ∧¬writing,
           Hole theResource.write[update]   // leave cheese while writing after
             after writing ≔ True          // recording that writing is happening
             afterward (IsEmpty readersQ) �
                         True ⦂ permit writersQ always writing ≔ False,
                           False ⦂ permit readersQ always writing ≔ False⍰§▮
```

Symbols
≡ → ⦂ ∧ ∨ ¬
⍰ ¶ § ▮

Illustration of writing-priority:

```
Actor WritingPriority[theResource:ReadersWriter]
   queues readersQ, writersQ,
   writing ≔ False,
   numberReading:(Integer thatIs ≧0) ≔ 0,
   implements ReadersWriter using
    read[query]→
     Do (writing ∨ ¬Empty writersQ) �
          True ⦂ Enqueue readersQ⬣ Void  // leave cheese while in readersQ
                      backout ¬writing ∧ numberReading=0 ∧ IsEmpty readersQ �
                                   True ⦂ Void permit writersQ,
                                   False ⦂ Void ⍰,
          False ⦂ Void ⍰⬣
        Precondition ¬writing,
            Hole theResource.read[query]
               after IsEmpty writersQ �
                      True ⦂ Permit readersQ always numberReading++,
                      False ⦂ Also numberReading++⍰
               afterward
                   (IsEmpty writersQ) �
                      True ⦂ permit readersQ always numberReading––,
                      False ⦂ numberReading �
                                 1 ⦂ permit writersQ always numberReading––
                                 else ⦂ numberReading––⍰ ⍰ ¶
    write[update]→
       Do numberReading>0 ∨ ¬IsEmpty readersQ ∨ writing ∨ ¬IsEmpty writersQ �
           True ⦂ Enqueue writersQ⬣ Void   // leave cheese while in writersQ
                       backout (IsEmpty writersQ ∧ ¬writing) �
                                 True ⦂ Void permit readersQ,
                                 False ⦂ Void ⍰,
           False ⦂ Void ⍰⬣
         Precondition numberReading=0 ∧ ¬writing,
            Hole theResource.write[update]
              after writing ≔ True
              afterward
                (IsEmpty readersQ) �
                   True ⦂ permit writersQ  always writing ≔ False,
                   False ⦂ permit readersQ always writing ≔ False⍰§▮
```

| Symbols |
|---|
| ≡ → ⦂ ∧ ∨ ¬ |
| ⍰ ¶ § ▮ |

The formal syntax of queue management in cheese is in the following end note: **49.**

## Conclusion

Before long, we will have billions of chips, each with hundreds of hyper-threaded cores executing hundreds of thousands of threads. Consequently, GOFIP (Good Old-Fashioned Imperative Programming) paradigm must be fundamentally extended. ActorScript is intended to be a contribution to this extension.

## Acknowledgements

Important contributions to the semantics of Actors have been made by: Gul Agha, Beppe Attardi, Henry Baker, Will Clinger, Irene Greif, Carl Manning, Ian Mason, Ugo Montanari, Maria Simi, Scott Smith, Carolyn Talcott, Prasanna Thati, and Aki Yonezawa.

Important contributions to the implementation of Actors have been made by: Bill Athas, Russ Atkinson, Beppe Attardi, Henry Baker, Gerry Barber, Peter Bishop, Nanette Boden, Jean-Pierre Briot, Bill Dally, Peter de Jong, Jessie Dedecker, Ken Kahn, Henry Lieberman, Carl Manning, Mark S. Miller, Tom Reinhardt, Chuck Seitz, Dale Schumacher, Richard Steiger, Dan Theriault, Mario Tokoro, Darrell Woelk, and Carlos Varela.

Research on the Actor model has been carried out at Caltech Computer Science, Kyoto University Tokoro Laboratory, MCC, MIT Artificial Intelligence Laboratory, SRI, Stanford University, University of Illinois at Urbana-Champaign Open Systems Laboratory, Pierre and Marie Curie University (University of Paris 6), University of Pisa, University of Tokyo Yonezawa Laboratory and elsewhere.

The members of the Silicon Valley Friday AM group made valuable suggestions for improving this paper. Discussions with Blaine Garst were helpful in the development of the implementation of Swiss cheese that doesn't hold a lock as well providing background on the historical development of interfaces. Patrick Beard found bugs and suggested improvements in presentation. Fanya S. Montalvo and Ike Nassi suggested simplifying the syntax. Dale Schumacher found many typos, suggested including a syntax diagram, and suggested improvements to the syntax of collections, binding and assignment. In particular, Dale contributed greatly to the development of the lock-free[i] implementation of cheese in the appendix. Chip Morningstar provided an excellent critique with many useful comments and suggestions.

ActorScript is intended to provide a foundation for information integration in privacy-friendly client-cloud computing [Hewitt 2009b].

---

[i] In the sense that the implementation never spins on a hardware lock.

# Bibliography

Hal Abelson and Gerry *Sussman Structure and Interpretation of Computer Programs* 1984.

Paul Abrahams. *A final solution to the Dangling else of ALGOL 60 and related languages* CACM. September 1966.

Sarita Adve and Hans-J. Boehm *Memory Models: A Case for Rethinking Parallel Languages and Hardware* CACM. August 2010.

Mikael Amborn. *Facet-Oriented Program Design*. LiTH-IDA-EX–04/047–SE Linkőpings Universitet. 2004.

Joe Armstrong *History of Erlang* HOPL III. 2007.

Joe Armstrong. *Erlang*. CACM. September 2010/

William Athas and Charles Seitz *Multicomputers: message-passing concurrent computers* IEEE Computer August 1988.

William Athas and Nanette Boden *Cantor: An Actor Programming System for Scientific Computing* in Proceedings of the NSF Workshop on Object-Based Concurrent Programming. 1988. Special Issue of SIGPLAN Notices.

Russ Atkinson. *Automatic Verification of Serializers* MIT Doctoral Dissertation. June, 1980.

Henry Baker. *Actor Systems for Real-Time Computation* MIT EECS Doctoral Dissertation. January 1978.

Henry Baker and Carl Hewitt *The Incremental Garbage Collection of Processes* Proceeding of the Symposium on Artificial Intelligence Programming Languages. SIGPLAN Notices 12, August 1977.

Paul Baran. *On Distributed Communications Networks* IEEE Transactions on Communications Systems. March 1964.

Gerry Barber. *Reasoning about Change in Knowledgeable Office Systems* MIT EECS Doctoral Dissertation. August 1981.

Philippe Besnard and Anthony Hunter. *Quasi-classical Logic: Non-trivializable classical reasoning from inconsistent information* Symbolic and Quantitative Approaches to Reasoning and Uncertainty. Springer LNCS. 1995.

Peter Bishop *Very Large Address Space Modularly Extensible Computer Systems* MIT EECS Doctoral Dissertation. June 1977.

Andreas Blass, Yuri Gurevich, Dean Rosenzweig, and Benjamin Rossman (2007a) *Interactive small-step algorithms I: Axiomatization* Logical Methods in Computer Science. 2007.

Andreas Blass, Yuri Gurevich, Dean Rosenzweig, and Benjamin Rossman (2007b*) Interactive small-step algorithms II: Abstract state machines and the characterization theorem.* Logical Methods in Computer Science. 2007.

Per Brinch Hansen *Monitors and Concurrent Pascal: A Personal History* CACM 1996.

Don Box, David Ehnebuske, Gopal Kakivaya, Andrew Layman, Noah Mendelsohn, Henrik Nielsen, Satish Thatte, Dave Winer. *Simple Object Access Protocol (SOAP) 1.1* W3C Note. May 2000.

Jean-Pierre Briot. *Acttalk: A framework for object-oriented concurrent programming-design and experience* 2nd France-Japan workshop. 1999.

Jean-Pierre Briot. *From objects to Actors: Study of a limited symbiosis in Smalltalk-80* Rapport de Recherche 88-58, RXF-LITP. Paris, France. September 1988.

Luca Cardelli, James Donahue, Lucille Glassman, Mick Jordan, Bill Kalsow, Greg Nelson. *Modula-3 report (revised)* DEC Systems Research Center Research Report 52. November 1989.

Luca Cardelli and Andrew Gordon *Mobile Ambients* FoSSaCS'98.

Arnaud Carayol, Daniel Hirschkoff, and Davide Sangiorgi. *On the representation of McCarthy's amb in the π-calculus* "Theoretical Computer Science" February 2005.

Alonzo Church "A Set of postulates for the foundation of logic (1&2)" Annals of Mathematics. Vol. 33, 1932. Vol. 34, 1933.

Alonzo Church *The Calculi of Lambda-Conversion* Princeton University Press. 1941.

Will Clinger. *Foundations of Actor Semantics* MIT Mathematics Doctoral Dissertation. June 1981.

Tyler Close *Web-key: Mashing with Permission* WWW'08.

Eric Crahen. Facet: A pattern for dynamic interfaces. CSE Dept. SUNY at Buffalo. July 22, 2002.

Haskell Curry and Robert Feys. *Combinatory Logic.* North-Holland. 1958.

Ole-Johan Dahl and Kristen Nygaard. "Class and subclass declarations" *IFIP TC2 Conference on Simulation Programming Languages.* 1967.

William Dally and Wills, D. *Universal mechanisms for concurrency* PARLE '89.

William Dally, et al. *The Message-Driven Processor: A Multicomputer Processing Node with Efficient Mechanisms* IEEE Micro. April 1992.

Jack Dennis and Earl Van Horn. *Programming Semantics for Multiprogrammed Computations* CACM. March 1966.

Edsger Dijkstra. *Cooperating sequential processes* Technical Report EWD-123, Technological University, Eindhoven, The Netherlands. 1965.

Edsger Dijkstra. *Go To Statement Considered Harmful* Letter to Editor CACM. March 1968.

Jason Eisner and Nathaniel W. Filardo. *Dyna: Extending Datalog for modern AI.* Datalog Reloaded. Springer. 2011.

Arthur Fine. *The Shaky Game: Einstein Realism and the Quantum Theory* University of Chicago Press, Chicago, 1986.

Frederic Fitch. *Symbolic Logic: an Introduction.* Ronald Press. 1952.

Nissim Francez, Tony Hoare, Daniel Lehmann, and Willem-Paul de Roever. *Semantics of nondeterminism, concurrency, and communication* Journal of Computer and System Sciences. December 1979.

Christopher Fuchs *Quantum mechanics as quantum information (and only a little more)* in A. Khrenikov (ed.) Quantum Theory: Reconstruction of Foundations (Växjo: Växjo University Press, 2002).

Blaine Garst. *Origin of Interfaces* Email to Carl Hewitt on October 2, 2009.

Elihu M. Gerson. *Prematurity and Social Worlds* in Prematurity in Scientific Discovery. University of California Press. 2002.

Andreas Glausch and Wolfgang Reisig. *Distributed Abstract State Machines and Their Expressive Power* Informatik Berichete 196. Humboldt University of Berlin. January 2006.

Brian Goetz [State of the Lambda] Brian Goetz's Oracle Blog. July 6, 2010.

Adele Goldberg and Alan Kay (ed.) *Smalltalk-72 Instruction Manual* SSL 76-6. Xerox PARC. March 1976.

Dina Goldin and Peter Wegner. *The Interactive Nature of Computing: Refuting the Strong Church-Turing Thesis* Minds and Machines March 2008.

Cordell Green. *Application of Theorem Proving to Problem Solving* IJCAI'69.
Irene Greif and Carl Hewitt. *Actor Semantics of PLANNER-73* Conference Record of ACM Symposium on Principles of Programming Languages. January 1975.
Irene Greif. *Semantics of Communicating Parallel Professes* MIT EECS Doctoral Dissertation. August 1975.
William Gropp, et. al. *MPI—The Complete Reference: Volume 2, The MPI-2 Extensions*. MIT Press. 1998
Pat Hayes *Some Problems and Non-Problems in Representation Theory* AISB. Sussex. July, 1974
Werner Heisenberg. *Physics and Beyond: Encounters and Conversations* translated by A. J. Pomerans (Harper & Row, New York, 1971), pp. 63 – 64.
Carl Hewitt. *More Comparative Schematology* MIT AI Memo 207. August 1970.
Carl Hewitt, Peter Bishop and Richard Steiger. *A Universal Modular Actor Formalism for Artificial Intelligence* IJCAI'73.
Carl Hewitt, *et al. Actor Induction and Meta-evaluation* Conference Record of ACM Symposium on Principles of Programming Languages, January 1974.
Carl Hewitt and Henry Lieberman. *Design Issues in Parallel Architecture for Artificial Intelligence* MIT AI memo 750. Nov. 1983.
Carl Hewitt, Tom Reinhardt, Gul Agha, and Giuseppe Attardi *Linguistic Support of Receptionists for Shared Resources* MIT AI Memo 781. Sept. 1984.
Carl Hewitt, *et al. Behavioral Semantics of Nonrecursive Control Structure* Proceedings of *Colloque sur la Programmation*, April 1974.
Carl Hewitt. *How to Use What You Know* IJCAI. September, 1975.
Carl Hewitt. *Viewing Control Structures as Patterns of Passing Messages* AI Memo 410. December 1976. Journal of Artificial Intelligence. June 1977.
Carl Hewitt and Henry Baker *Laws for Communicating Parallel Processes* IFIP-77, August 1977.
Carl Hewitt and Russ Atkinson. *Specification and Proof Techniques for Serializers* IEEE Journal on Software Engineering. January 1979.
Carl Hewitt, Beppe Attardi, and Henry Lieberman. *Delegation in Message Passing* Proceedings of First International Conference on Distributed Systems Huntsville, AL. October 1979.
Carl Hewitt and Gul Agha. *Guarded Horn clause languages: are they deductive and Logical?* in Artificial Intelligence at MIT, Vol. 2. MIT Press 1991.
Carl Hewitt and Jeff Inman. *DAI Betwixt and Between: From "Intelligent Agents" to Open Systems Science* IEEE Transactions on Systems, Man, and Cybernetics. Nov./Dec. 1991.
Carl Hewitt and Peter de Jong. *Analyzing the Roles of Descriptions and Actions in Open Systems* Proceedings of the National Conference on Artificial Intelligence. August 1983.
Carl Hewitt. (2006). "What is Commitment? Physical, Organizational, and Social" *COIN@AAMAS'06*. (Revised version to be published in Springer Verlag Lecture Notes in Artificial Intelligence. Edited by Javier Vázquez-Salceda and Pablo Noriega. 2007) April 2006.
Carl Hewitt (2007a). "Organizational Computing Requires Unstratified Paraconsistency and Reflection" *COIN@AAMAS*. 2007.
Carl Hewitt (2008a) *Norms and Commitment for iOrgs*™ Information Systems: *Direct Logic*™ *and Participatory Argument Checking* ArXiv 0906.2756.

Carl Hewitt (2008b) “Large-scale Organizational Computing requires Unstratified Reflection and Strong Paraconsistency” *Coordination, Organizations, Institutions, and Norms in Agent Systems III* Jaime Sichman, Pablo Noriega, Julian Padget and Sascha Ossowski (ed.). Springer-Verlag. *http://organizational.carlhewitt.info/*

Carl Hewitt (2008e). *ORGs for Scalable, Robust, Privacy-Friendly Client Cloud Computing* IEEE Internet Computing September/October 2008.

Carl Hewitt (2008f) *Common sense for concurrency and inconsistency robustness using Direct Logic*™ and the Actor Model ArXiv 0812.4852.

Carl Hewitt (2009a) *Perfect Disruption: The Paradigm Shift from Mental Agents to ORGs* IEEE Internet Computing. Jan/Feb 2009.

Carl Hewitt (2009b) *A historical perspective on developing foundations for client-cloud computing: iConsult™ & iEntertain™ Apps using iInfo™ Information Integration for iOrgs™ Information Systems* (Revised version of “Development of Logic Programming: What went wrong, What was done about it, and What it might mean for the future” AAAI Workshop on What Went Wrong. AAAI-08.) ArXiv 0901.4934.

Carl Hewitt (2013) *Inconsistency Robustness in Logic Programs* ArXiv 0904.3036

Carl Hewitt (2010a) *Actor Model of Computation* arXiv:1008.1459

Carl Hewitt (2010b) *iTooling™: Infrastructure for iAdaptive™ Concurrency*

Carl Hewitt (editor). *Inconsistency Robustness 1011* Stanford University. 2011.

Carl Hewitt, Erik Meijer, and Clemens Szyperski “The Actor Model (everything you wanted to know, but were afraid to ask)” http://channel9.msdn.com/Shows/Going+Deep/Hewitt-Meijer-and-Szyperski-The-Actor-Model-everything-you-wanted-to-know-but-were-afraid-to-ask Microsoft Channel 9. April 9, 2012.

Carl Hewitt. *“Health Information Systems Technologies” http://ee380.stanford.edu/cgi-bin/videologger.php?target=120606-ee380-300.asx*
Slides for this video: *http://HIST.carlhewitt.info* Stanford CS Colloquium. June 6, 2012.

Carl Hewitt. *What is computation? Actor Model versus Turing's Model* in “A Computable Universe: Understanding Computation & Exploring Nature as Computation”. edited by Hector Zenil. World Scientific Publishing Company. 2012.

Tony Hoare *Quick sort* Computer Journal 5 (1) 1962.

Tony Hoare *Monitors: An Operating System Structuring Concept* CACM. October 1974.

Tony Hoare. *Communicating sequential processes* CACM. August 1978.

Tony Hoare. *Communicating Sequential Processes* Prentice Hall. 1985.

Tony Hoare. *Null References: The Billion Dollar Mistake.* QCon. August 25, 2009.

W. Horwat, Andrew Chien, and William Dally. *Experience with CST: Programming and Implementation* PLDI. 1989.

Anthony Hunter. *Reasoning with Contradictory Information using Quasi-classical Logic* Journal of Logic and Computation. Vol. 10 No. 5. 2000.

M. Jammer *The EPR Problem in Its Historical Development* in Symposium on the Foundations of Modern Physics: 50 years of the Einstein-Podolsky-Rosen Gedankenexperiment, edited by P. Lahti and P. Mittelstaedt. World Scientific. Singapore. 1985.

Simon Peyton Jones, Andrew Gordon, Sigbjorn Finne. *Concurrent Haskell*, POPL’96.

Ken Kahn. *A Computational Theory of Animation* MIT EECS Doctoral Dissertation. August 1979.

Alan Kay. "Personal Computing" in *Meeting on 20 Years of Computing Science* Instituto di Elaborazione della Informazione, Pisa, Italy. 1975. *http://www.mprove.de/diplom/gui/Kay75.pdf*

Frederick Knabe *A Distributed Protocol for Channel-Based Communication with Choice* PARLE'92.

Bill Kornfeld and Carl Hewitt. *The Scientific Community Metaphor* IEEE Transactions on Systems, Man, and Cybernetics. January 1981.

Bill Kornfeld. *Parallelism in Problem Solving* MIT EECS Doctoral Dissertation. August 1981.

Robert Kowalski. *A proof procedure using connection graphs* JACM. October 1975.

Robert Kowalski *Algorithm = Logic + Control* CACM. July 1979.

Robert Kowalski. *Response to questionnaire* Special Issue on Knowledge Representation. SIGART Newsletter. February 1980.

Robert Kowalski (1988a) *The Early Years of Logic Programming* CACM. January 1988.

Robert Kowalski (1988b) *Logic-based Open Systems* Representation and Reasoning. Stuttgart Conference Workshop on Discourse Representation, Dialogue tableaux and Logic Programming. 1988.

Edya Ladan-Mozes and Nir Shavit. *An Optimistic Approach to Lock-Free FIFO Queues* Distributed Computing. Sprinter. 2004.

Leslie Lamport *How to make a multiprocessor computer that correctly executes multiprocess programs* IEEE Transactions on Computers. 1979.

Peter Landin. *A Generalization of Jumps and Labels* UNIVAC Systems Programming Research Report. August 1965. (Reprinted in *Higher Order and Symbolic Computation.* 1998)

Peter Landin *A correspondence between ALGOL 60 and Church's lambda notation* CACM. August 1965.

Edward Lee and Stephen Neuendorffer *Classes and Subclasses in Actor-Oriented Design*. Conference on Formal Methods and Models for Codesign (MEMOCODE). June 2004.

Steven Levy *Hackers: Heroes of the Computer Revolution* Doubleday. 1984.

Henry Lieberman. *An Object-Oriented Simulator for the Apiary* Conference of the American Association for Artificial Intelligence, Washington, D. C., August 1983

Henry Lieberman. *Thinking About Lots of Things at Once without Getting Confused: Parallelism in Act 1* MIT AI memo 626. May 1981.

Henry Lieberman. *A Preview of Act 1* MIT AI memo 625. June 1981.

Henry Lieberman and Carl Hewitt. *A real Time Garbage Collector Based on the Lifetimes of Objects* CACM June 1983.

Barbara Liskov and Liuba Shrira *Promises: Linguistic Support for Efficient Asynchronous Procedure Calls* SIGPLAN'88.

Barbara Liskov and Jeannette Wing . *A behavioral notion of subtyping*, TOPLAS, November 1994.

Carl Manning. *Traveler: the Actor observatory* ECOOP 1987. Also appears in Lecture Notes in Computer Science, vol. 276.

Carl Manning. *Acore: The Design of a Core Actor Language and its Compile* Master Thesis. MIT EECS. May 1987.

Satoshi Matsuoka and Aki Yonezawa. *Analysis of Inheritance Anomaly in Object-Oriented Concurrent Programming Languages Research Directions in Concurrent Object-Oriented Programming* MIT Press. 1993.

John McCarthy *Programs with common sense* Symposium on Mechanization of Thought Processes. National Physical Laboratory, UK. Teddington, England. 1958.

John McCarthy. *A Basis for a Mathematical Theory of Computation* Western Joint Computer Conference. 1961.

John McCarthy, Paul Abrahams, Daniel Edwards, Timothy Hart, and Michael Levin. *Lisp 1.5 Programmer's Manual* MIT Computation Center and Research Laboratory of Electronics. 1962.

John McCarthy. *Situations, actions and causal laws* Technical Report Memo 2, Stanford University Artificial Intelligence Laboratory. 1963.

John McCarthy and Patrick Hayes. *Some Philosophical Problems from the Standpoint of Artificial Intelligence* Machine Intelligence 4. Edinburgh University Press. 1969.

Alexandre Miquel. *A strongly normalising Curry-Howard correspondence for IZF set theory* in Computer science Logic Springer. 2003

Giuseppe Milicia and Vladimiro Sassone. *The Inheritance Anomaly: Ten Years After* SAC. Nicosia, Cyprus. March 2004.

Mark S. Miller *Robust Composition: Towards a Unified Approach to Access Control and Concurrency Control* Doctoral Dissertation. John Hopkins. 2006.

Mark S. Miller *et. al. Bringing Object-orientation to Security Programming*. YouTube. November 3, 2011.

George Milne and Robin Milner. "Concurrent processes and their syntax" *JACM*. April, 1979.

Robert Milne and Christopher Strachey. *A Theory of Programming Language Semantics* Chapman and Hall. 1976.

Robin Milner. *Logic for Computable Functions: description of a machine implementation.* Stanford AI Memo 169. May 1972

Robin Milner *Processes: A Mathematical Model of Computing Agents* Proceedings of Bristol Logic Colloquium. 1973.

Robin Milner *Elements of interaction: Turing award lecture* CACM. January 1993.

Marvin Minsky (ed.) *Semantic Information Processing* MIT Press. 1968.

Eugenio Moggi *Computational lambda-calculus and monads* IEEE Symposium on Logic in Computer Science. Asilomar, California, June 1989.

Allen Newell and Herbert Simon. *The Logic Theory Machine: A Complex Information Processing System.* Rand Technical Report P-868. June 15, 1956

Carl Petri. *Kommunikation mit Automate* Ph. D. Thesis. University of Bonn. 1962.

Simon Peyton Jones, Alastair Reid, Fergus Henderson, Tony Hoare, and Simon Marlow. *A semantics for imprecise exceptions* Conference on Programming Language Design and Implementation. 1999.

Paul Philips. *We're Doing It all Wrong* Pacific Northwest Scala 2013.

Gordon Plotkin. *A powerdomain construction* SIAM Journal of Computing. September 1976.

George Polya (1957) *Mathematical Discovery: On Understanding, Learning and Teaching Problem Solving Combined Edition* Wiley. 1981.

Karl Popper (1935, 1963) *Conjectures and Refutations: The Growth of Scientific Knowledge* Routledge. 2002.

John Reppy, Claudio Russo, and Yingqi Xiao *Parallel Concurrent ML* ICFP'09.

John Reynolds. *Definitional interpreters for higher order programming languages* ACM Conference Proceedings. 1972.
Bill Roscoe. *The Theory and Practice of Concurrency* Prentice-Hall. Revised 2005.
Kenneth Ross, Yehoshua Sagiv. *Monotonic aggregation in deductive databases.* Principles of Distributed Systems. June 1992Dana Scott and Christopher Strachey. *Toward a mathematical semantics for computer languages* Oxford Programming Research Group Technical Monograph. PRG-6. 1971
Charles Seitz. *The Cosmic Cube* CACM. Jan. 1985.
Peter Sewell, et. al. *x86-TSO: A Rigorous and Usable Programmer's Model for x86 Microprocessors* CACM. July 2010.
Michael Smyth. *Power domains* Journal of Computer and System Sciences. 1978.
Guy Steele, Jr. *Lambda: The Ultimate Declarative* MIT AI Memo 379. November 1976.
Guy Steele, Jr. *Debunking the 'Expensive Procedure Call' Myth, or, Procedure Call Implementations Considered Harmful, or, Lambda: The Ultimate GOTO.* MIT AI Lab Memo 443. October 1977.
Gunther Stent. *Prematurity and Uniqueness in Scientific Discovery* Scientific American. December, 1972.
Bjarne Stroustrup *Programming Languages — C++* ISO N2800. October 10, 2008.
Gerry Sussman and Guy Steele *Scheme: An Interpreter for Extended Lambda Calculus* AI Memo 349. December, 1975.
David Taenzer, Murthy Ganti, and Sunil Podar, *Problems in Object-Oriented Software Reuse* ECOOP'89.
Daniel Theriault. *A Primer for the Act-1 Language* MIT AI memo 672. April 1982.
Daniel Theriault. *Issues in the Design and Implementation of Act 2* MIT AI technical report 728. June 1983.
Hayo Thielecke *An Introduction to Landin's "A Generalization of Jumps and Labels"* Higher-Order and Symbolic Computation. 1998.
Dave Thomas and Brian Barry. *Using Active Objects for Structuring Service Oriented Architectures: Anthropomorphic Programming with Actors* Journal of Object Technology. July-August 2004.
Kazunori Ueda *A Pure Meta-Interpreter for Flat GHC, A Concurrent Constraint Language* Computational Logic: Logic Programming and Beyond. Springer. 2002.
Darrell Woelk. *Developing InfoSleuth Agents Using Rosette: An Actor Based Language* Proceedings of the CIKM '95 Workshop on Intelligent Information Agents. 1995.
Akinori Yonezawa, Ed. *ABCL: An Object-Oriented Concurrent System* MIT Press. 1990.
Aki Yonezawa *Specification and Verification Techniques for Parallel Programs Based on Message Passing Semantics* MIT EECS Doctoral Dissertation. December 1977.
Hadasa Zuckerman and Joshua Lederberg. *Postmature Scientific Discovery?* Nature. December, 1986.

## Appendix 1. Extreme ActorScript

**Parameterized Types,** *i.e.,* **◁ , ▷**

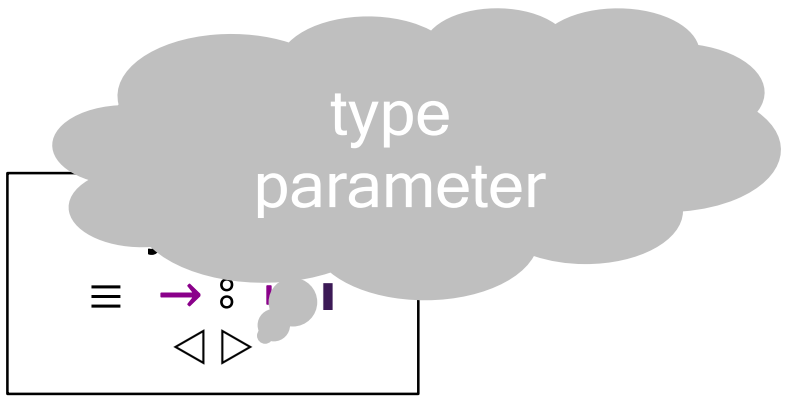

Parameterized Types are specialized using other types delimited by "◁" and "▷":

Double◁aType▷ ≡
    **Actor implements** [aType]↦aType **using**
        [x]→ x+x §▮    // addition for aType

The formal syntax of parameterized types is in the following end note: **50 .**

**Structures, i.e., Structure**

A structure can be defined using aa structure identifier followed a list of the parts enclosed by "[" and "]".

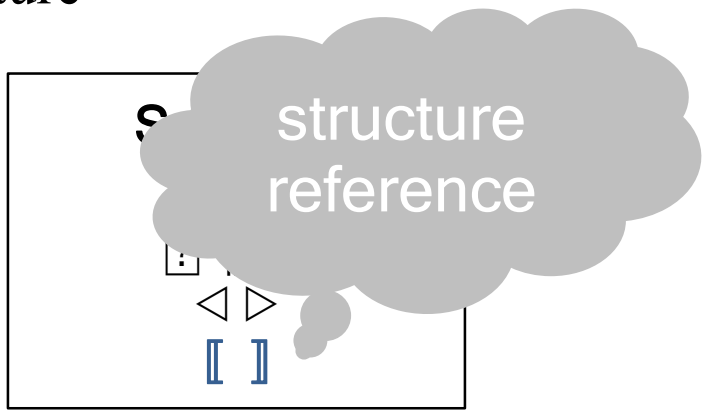

For example, the structure Leaf can be defined as follows to extend Tree:

**Structure** Leaf◁aType▷[aTerminal:aType] **extends** Tree◁aType▷▮
    // a terminal must be of type aType

For example,

- The expression **Let** x[i] ← 3, Leaf◁Integer▷[x]▮ is equivalent to Leaf◁Integer▷[3]▮
- The pattern Leaf◁Integer▷[x] matches Leaf◁Integer▷[3] and binds x to 3.

The formal syntax of structures is in the following end note: **51**

**Structures with named fields, i.e., ⌸ and :⌸**

The structure Fork can be defined as follows:

**Structure** Fork◁aType▷[left⌸ Tree, right⌸ Tree]
    **extends** Tree◁aType▷▮

---

[i] x is of type Integer

For example,

- The expression
  Let x ← 3, Fork◁Integer▷[left⊟ Leaf◁Integer▷[x],
  right⊟ Leaf◁Integer▷[ x+1]])▮
  is equivalent to the following:
  Fork◁Integer▷[Leaf◁Integer▷[left⊟ 3],
  right⊟ Leaf◁Integer▷[4]]▮
- The pattern Fork◁Integer▷[left⊟ x, right⊟ y] matches Fork◁Integer▷[Leaf◁Integer▷[6], Leaf◁Integer▷[6]] and binds x to Leaf◁Integer▷[5] and y to Leaf◁Integer▷[6].

The formal syntax structures with named fields is in the following end note: **52.**

**Processing Exceptions,** *i.e.,* **Try catch�** ⁸ **,** ⁸ ⍰ *and* **Try cleanup**

It is useful to be able to catch exceptions. The following illustration returns the string "This is a test.":

Try Throw Exception["This is a test."] catch�
Exception[aString:String] ⁸ aString ⍰▮

The following illustration performs Reset.[ ] and then rethrows Exception["This is another test."]:

Try Throw Exception["This is another test."] cleanup Reset.[ ]▮

The formal syntax of processing exceptions is in the following end note: **53.**

**Runtime Requirements,** *i.e.,* **Precondition ; and postcondition**

A runtime requirement throws exception an exception if does not hold.
For example, the following expression throws an exception that the requirement x≥0 doesn't hold:

Let x ←−1,
Precondition x≥0,
SquareRoot.[x]▮

Post conditions can be tested using a procedure. For example, the following expression throws an exception that **postcondition** failed because square root of 2 is not less than 1:

SquareRoot.[2] postcondition [y:Float]→ y<1▮

The formal syntax requirements is in the following end note: **54**.

**Polymorphism**

Polymorphism provides for multiple implementations of a type. For example, **Cartesian** Actors that implement **Complex**[i] can be defined as follows:

```
Actor Cartesian[myReal:Float default 0, myImaginary:Float default 0]
  implements Complex using      // construct a Cartesian of type Complex
    realPart[ ]→ myReal¶
    imaginaryPart[ ]→ myImaginary¶
    magnitude[ ]→
         SquareRoot.[myReal*myReal + myImaginary*myImaginary]¶
    angle[ ]→
      Let theta ← Arcsine.[myImaginary/..magnitude[ ]],
            // ..magnitude[ ] is the result of sending magnitude[ ] to this Actor
         myReal>0 �
            True ⦂ theta,
            False ⦂ myImaginary >0 �
                          True ⦂180º−theta,55
                          False ⦂180º+theta ⍰ ⍰¶
    plus[argument]→
      Let argumentRealPart ← argument.realPart[ ],
           argumentImaginaryPart ← argument.imaginaryPart[ ],
        Cartesian.[myReal+argumentRealPart,
                    myImaginary+argumentImaginaryPart]¶
    times[argument]→
      Let {argumentRealPart ← argument.realPart[ ],
            argumentImaginaryPart ← argument.imaginaryPart[ ]},
        Cartesian.[myReal*argumentRealPart
                      – myImaginary*argumentImaginaryPart,
                     myImaginary*argumentRealPart
                        + myReal*argumentImaginaryPart]¶
    equivalent[x] →                  //  test if x is an equivalent complex number
      myReal=z.realPart[ ] ∧ myImaginary=z.imaginaryPart[ ]§▮
```

---

[i] Interface Complex {realPart[ ]↦ Float,
imaginaryPart[ ]↦ Float,
magnitude[ ]↦ Float,
angle[ ]↦ Degrees,
plus[Complex]↦ Complex,
times[Complex]↦ Complex,
equivalent[Complex]↦ Boolean}▮

Consequently,

- Cartesian.[1, 2].realPart[ ]▮ is equivalent to 1▮
- Cartesian.[3, 4].magnitude[ ]▮ is equivalent to 5.0▮
- Cartesian.[0, 1].times[Cartesian.[0, 1]]▮ is equivalent to Cartesian.[-1, 0]▮[56]
- Cartesian.[1, 2]:Complex▮ is equivalent to True▮
- Cartesian.[1, 2]:Cartesian▮ is equivalent to False▮ because the constructor returns Actors of type Complex

**Arguments with named fields, *i.e.*, ⌸ and :⌸**

Polar Actors that implement Complex with named arguments angle and magnitude can be defined as follows:

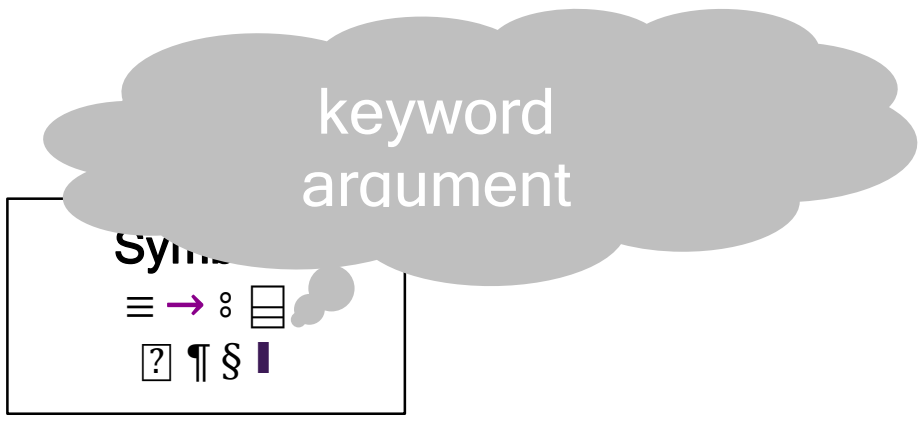

Actor Polar[angle ⌸ Degrees **default** 0°,
// angle of type Degrees is a named argument of Polar with
// default 0°
magnitude ⌸ Length]
**implements** Complex **using**
angle[ ]→ angle¶

realPart[ ]→ magnitude∗Sine.[angle]¶
imaginaryPart[ ]→ magnitude∗Cosine.[angle]¶
plus[argument]→
Cartesian.[argument.realPart[ ] + ..realPart[ ],
// ..realPart [ ] is the result of sending realPart [ ] to this Actor
argument.imaginaryPart[ ] + ..imaginaryPart[ ]]¶
times[argument]→
Polar.[angle⌸ angle+argument.angle[ ],
magnitude⌸ magnitude∗argument.magnitude[ ]]¶
equivalent[x]→
x � z:Complex ⦂ ..realPart[ ]=z.realPart[ ]
∧ ..imaginaryPart[ ]=z.imaginaryPart[ ],
else ⦂ False ⍰§▮

Consequently,

- Polar.[theAngle ⌸ 0°, theMagnitude ⌸ 1].realPart[ ]▮ is equivalent to 1
- (Polar.[theMagnitude ⌸ 1]).equivalent[Cartesian.[1, 0]]▮ is equivalent to True▮

**Lists,** *i.e.,* **[ ] using Spread,** *i.e.,* **[ ⩛ ]**

A list expression begins with “List” followed by the type of list element[i] and expressions for list elements[ii]. Similarly “Lists” is used for a list of lists. The prefix operator "⩛" can be sued to spread the elements of a list. For example

- List◁Integer▷[1, ⩛[2, 3], 4]▮ is equivalent to List◁Integer▷[1, 2, 3, 4]▮.
- Lists◁Integer▷[[1, 2], ⩛[3, 4]]▮ is equivalent to Lists◁Integer[[1, 2], 3, 4]▮
- If y is List◁Integer▷[5, 6], then Lists◁Integer▷[1, 2, ⩛[y], ⩛y]▮ is equivalent to Lists◁Integer▷[1, 2, [5, 6], 5, 6]▮
- List◁Integer▷[1, 2] is the list of integers of type Integer with just 1 and 2.
- List◁Integer▷[1, 2.0] throws an exception because 2.0 is not of type Integer

The formal syntax of list expressions is in the following end note: **57.**

A list pattern begins with “List” followed by the type of list element[iii] and patterns for list elements[iv]. Within a list, “⩛”is used to match the pattern that follows with the list zero or more elements. Similarly “Lists” is used for a list of lists. For example:

- Lists◁Integer▷[[x, 2], ⩛y] is a pattern that matches [[1, 2], 3, 4] and binds x to 1 and y to [3, 4]
- Lists◁Integer▷[[1, 2], ⩛$$y] is a pattern that only matches Lists◁Integer▷[[1, 2], 3, 4] if y is Lists◁Integer▷[3, 4]
- List◁Integer▷[⩛x, ⩛y] is an illegal pattern because it can match ambiguously

The formal syntax of patterns is in the following end note: **58.**

---

[i] delimited by ◁ and ▷
[ii] delimited by “[” and “]”
[iii] delimited by ◁ and ▷
[iv] delimited by “[” and “]”

As an example of the use of spread, the following procedure returns every other element of a list beginning with the first:[59]

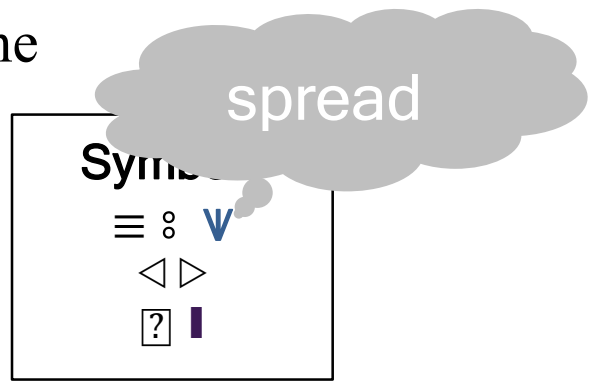

```
AlternateElements.[aList:List◁aType▷]:List◁aType▷ ≡
 aList �
  List◁aType▷[ ] ⦂ [ ],
  List◁aType▷[anElement] ⦂ [anElement],
  List◁aType▷[firstElement, secondElement] ⦂ [firstElement],
  else
   List◁aType▷[firstElement, secondElement, ∨remainingElements] ⦂
        [firstElement, ∨AlternateElements.[remainingElements]] ⍰▮
```

Consequently,

- AlternateElements.[List◁Integer▷[ ]]▮ is equivalent to List◁Integer▷[ ]▮
- AlternateElements.[List◁Integer▷[3]]▮ is equivalent to List◁Integer▷[3]]▮
- AlternateElements.[List◁Integer▷[3, 4]]▮ is equivalent to List◁Integer▷[3]▮
- AlternateElements.[List◁Integer▷[3, 4, 5]]▮ is equivalent to List◁Integer▷[3, 5]▮

**Sets,** *i.e.*, { } **using spreading,** *i.e.***,** { ∨ }

A set is an unordered structure with duplicates removed.

The formal syntax of sets is in the following end note: **60.**

**Multisets,** *i.e.*, {| |} **using spreading,** *i.e.*, {| ∨ |}

A set is an unordered structure with duplicates allowed.

The formal syntax of multisets is in the following end note: **61.**

**Maps,** *i.e.*, **Map**{ }

A map is composed of pairs. For example Map{ [3, “a”], [“x”, “b”]}▮

Pairs in maps are unordered, *e.g.*, Map{[3, “a”], [“x”, “b”]}▮ is equivalent to Map{[“x”, “b”], [3, “a”]}▮.

However, the expression Map{[“y”, “b”], [“y”, “a”]} throws an exception because a map is univalent. As another example, for the contact records of 1.1 billion people, the following can compute a list of pairs from age to average number of social contacts of US citizens sorted by increasing age:

```
Age ≡ Integer thatIs ≧0≦130▮

AgeToAverageOfNumberOfContactsPairsSortedByAge
          .[records:Set◁ContactRecord[i]▷]:List◁[Age, Float]▷ ≡[62]
  records.filter[ii][[aRecord:ContactRecord] determinate →
                    aRecord⟦citizenship⟧ �
                       "US" ⦂ True,
                       else ⦂ False ⍰]
          .collect[iii][[aRecord:ContactRecord] determinate →
               [aRecord⟦yearsOld⟧,
                aRecord⟦numberOfContacts⟧]]
          .reduceRange[iv]
             [[aSetOfNumberOfContacts:Set◁Integer▷] determinate →
                aSetOfNumberOfContacts.average[v][ ]]
          .sort[vi][LessThanOrEqual]▮
```

The formal syntax of maps is in the following end note: **63.**

---

[i] Structure ContactRecord[yearsOld ⌸ Age,
numberOfContacts ⌸ Integer,
citizenship ⌸ String]▮

[ii] Set◁ContactRecord▷ has filter[[ContactRecord]↦Boolean]↦
Set◁ContactRecord▷▮

[iii] Set◁ContactRecord▷ has collect [[ContactRecord]↦ [Age, Integer]]↦
Map◁Age, Set◁Integer▷▷]▮

[iv] Map◁Age, Set◁Integer▷▷ has reduceRange[[◁Set◁Integer▷]↦Float]↦
Map◁Age, Float▷▮

[v] Set◁Number▷ has [average[ ]↦ Float▮

[vi] Map◁Age, Float▷ has sort[[Age, Age]↦ Boolean)]↦ List◁[Age, Float]▷▮

**Futures, *i.e.*, Future** and **↓**

A future [Baker and Hewitt 1977] for an expression can be created in ActorScript by using "**Future**" preceding the expression. The operator "↓" can be used to "resolve" a future by returning an Actor computed by the future or throwing an exception. For example, the following expression is equivalent to Factorial.[9999]▮

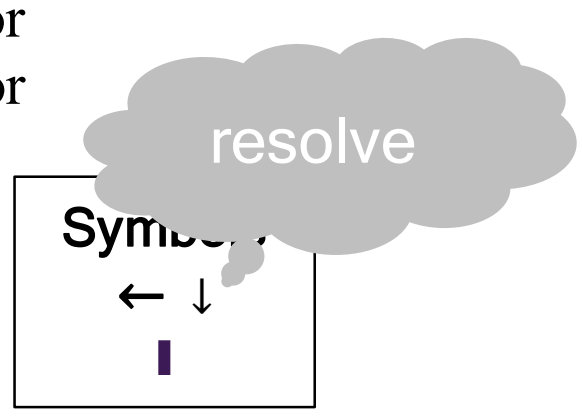

```
Let aFuture[i] ←Future Factorial.[9999],
  ↓aFuture▮     // do not proceed until Factorial.[9999] has
                 //  resolved[ii]
```

Futures allow execution of expressions to be adaptively executed indefinitely into the future.[64] For example, the following returns a future

```
Let aFuture ←Future Factorial.[9999],
    g ← ([afuture:Future◁Integer▷]→ 5),
                              // g returns 5 regardless of its argument
   g.[aFuture])▮
       // return 5 regardless of whether Factorial.[9999] has completed[iii]
```

Note that the following are all equivalent[65]:

- ↓**Future** (4+Factorial.[9999])▮
- 4+↓**Future** Factorial.[9999]▮
- 4+⦷Factorial.[9999]▮
- ⦷(4+Factorial.[9999])▮

Also ⦷Factorial.[9999]+⦷Fibonacci.[9000]▮ is equivalent to the following:

```
Let {n[iv] ←⦷Factorial.[9999],
     m ←⦷Fibonacci.[9000]}
   n+m▮   // return Factorial.[9999]+Fibonacci.[9000]
```

---

[i] f is of type Future◁Integer▷

[ii] i.e. returned or threw an exception

[iii] *i.e.* Factorial.[1000] might not have returned or thrown an exception when 5 is returned. The future f will be garbage collected.

[iv] n is of type Integer

In the following example, Factorial▪[9999] might never be executed if readCharacter▪[ ] returns the character 'x':

```
Let aFuture ← Future Factorial▪[9999],
  readCharacter▪[ ] �
      'x' ⦂ 1,                          // readCharacter▪[ ] returned 'x'
      else ⦂ 1+ ↓aFuture ⍰▮
                   // readCharacter▪[ ] returned something other than 'x'
```

In the above, program resolution of aFuture is highlighted in yellow.

Futures can be chained, as in the following example:

```
Size▪[aFutureList:Future◁List◁String▷▷]:Future◁Integer▷ ≡
  aFutureList �
    Future List◁String▷[ ] ⦂
      Future 0,
    Future List◁String▷[aFirst:String,
                        ⱯaRest:Future◁List◁String▷▷] ⦂
      Future aFirst▪length[ ] + Size▪[aRest] ⍰▮
```

The above procedure can computer the size of a list concurrently with creating the list. It does so by making use of a **partially resolved future** highlighted in yellow above.

Below is the definition of a procedure that computes a future of a list that is the "fringe" of the leaves of tree.[i]

```
Fringe▪[aTree:Tree◁aType▷]:Future◁List◁aType▷▷ ≡
    aTree � Leaf◁aType▷[x] ⦂ Future [x],
            Fork◁aType▷[tree1, tree2] ⦂
              Future
                [ⱯFringe▪[tree1], ⱯPostpone66 Fringe▪[tree2]] ⍰▮
```

[i] See definition of Tree above in this article.

The above procedure can be used to define SameFringe that determines if two lists have the same fringe [Hewitt 1972]:

```
SameFringe◁aType▷
          .[aTree:Tree◁aType▷, anotherTree:Tree◁aType▷]:Boolean ≡
                                        //  test if two trees have the same fringe
  Fringe.[aTree] �
     Future List◁aType▷[ ] ⦂
        Fringe.[aTree] �
           Future List◁aType▷[ ] ⦂ True,
           else ⦂ False
     Future List◁aType▷[first, ⩛rest] ⦂
        Fringe.[anotherTree] �
           Future List◁aType▷ [ ] ⦂ False,
           Future List◁aType▷[anotherFirst, ⩛anotherRest] ⦂
               first=anotherFirst �
                  True ⦂ SameFringe◁aType▷.[aRest, anotherRest],
                  False ⦂ False
```

The procedure below given a list of futures returns a future list with the same elements:

```
Futurize◁aType▷
   .[aListOfFutures:List◁Future◁aType▷▷]:Future◁List◁aType▷▷ ≡
   aListOfFutures �
     List◁Future◁aType▷▷[ ] ⦂
        Future List◁aType▷[ ],
     List◁Future◁aType▷▷[aFirst:Future◁aType▷,
                              ⩛aRest:List◁Future◁aType▷▷] ⦂
        Future List◁aType▷[↓aFirst,
                             ⩛Futurize◁aType▷.[aRest]] ▮
```

The formal syntax of futures is in the following end note: **67.**

**In-line Recursion (*e.g.*, looping) , *i.e.* ▪[ ← , ← ] ≜**

Inline recursion (often called looping) is accomplished using an initial invocation with identifiers initialized using "←" followed by "≜" and the body.[i]

Below is an illustration of a loop Factorial with two loop identifiers n and accumulation. The loop starts with n equals 9 and value equal 1. The loop is iterated by a call to Factorial with the loop identifiers as arguments.

Factorial▪[n ←9, accumulation ←1] ≜
  n � 1 ⦂ accumulation,
    **else** ⦂ Factorial▪[n–1, n∗ accumulation] ▮[ii]

The above compiles as a loop because the call to Factorial in the body is a "tail call" [Hewitt 1970, 1976; Steele 1977].

The following returns a list of ten times successively calling the procedure P[iii] (of type [ ]↦**Integer**) in order with no arguments:

FirstTenSequentially▪[n ←10]:**List**◁**Integer**▷ ≜
  n � 1 ⦂ [P▪[ ]],
    **else** ⦂ **Let** x ← P▪[ ];
      [x, ⩛FirstTenSequentially▪[n–1]] ▮

The following returns one of the results of concurrently calling the procedure P[iv] (of type [ ]↦**Integer**) ten times with no arguments:

OneOfTen▪[n ←10]:**List**◁**Integer**▷ ≜
  n � 1 ⦂ P▪[ ],
    **else** ⦂ ⓘP▪[ ] **either** ⓘOneOfTen▪[n–1]] ▮

The formal syntax of looping is in the following end note: **68.**

**Type Discrimination, *i.e.*, Discrimination and Δ**

A discrimination is a type of alternatives differentiated by type using "**Discrimination**" followed by a type name, "**{**", a list of types separated using

---

[i] This construct takes the place of **while**, **for**, *etc.* loops used in other programming languages.

[ii] equivalent to the following:

Factorial▪[n:**Integer** ←9, accumulation:**Integer** ←1]:**Integer** ≜
  n � 1 ⦂ accumulation,
    **else** ⦂ Factorial▪[n–1, n∗ accumulation] ▮

[iii] The procedure P may be indeterminate, *i.e.*, return different results on successive calls.

[iv] The procedure P may be indeterminate, *i.e.*, return different results on different calls.

"**,**", and terminated by "**}**". A discriminate can be selected by using a discrimination followed by "Δ" and the type to be selected.

For example, consider the following definition:

**Discrimination** IntegerOrFloat {Integer, Float}▮

Consequently,

- (IntegerOrFloat◁Integer▷[3]) ΔInteger▮ is equivalent to 3▮.
- (IntegerOrFloat◁Float▷[3.0]) ΔInteger▮ throws an exception because Integer is not the same as the discriminant Float.
- The pattern x:Float matches IntegerOrFloat◁Float▷[3.0] and binds x to 3.0.
- The expression (IntegerOrFloat◁Float▷[3.0]):Float▮ is equivalent to **True**▮.
- IntegerOrFloat◁Float▷[3.0] � y:Integer ⦂ y-1, x:Float ⦂ x+1 ⍰▮ is equivalent to 4.0▮.

A nullable is a discrimination:

**Discrimination** Nullable◁aType▷ {aType, Null◁aType▷}▮

There is exactly one Actor that is of type Null◁*aType*▷, namely **Null** *aType*.

A nullable can be created as follows:

**Nullable** x:aType ≡ Nullable◁aType▷[x]

Basic (whose is understood by the pattern matcher) can be defined as follows:

**Discrimination** Basic
{Integer, Float, Character, Boolean, String,
Nullable◁Basic▷, List◁Basic▷, Set◁Basic▷,
Multiset◁Basic▷, Map◁Basic, Basic▷}▮

The formal syntax of type discrimination is in the following end note: **69.**

**Strings**

Strings are Actors that can be expressed using "String", "[", string arguments, and "]". For example,

- String['1', "23", '4']▮ is equivalent to "1234"▮.
- String['1', '2', "34", "56"]▮ is equivalent to "123456"▮.
- String[String['1', '2'], "34"]▮ is equivalent to "1234"▮.
- String[ ]▮ is equivalent to ""▮.

String patterns are delimited by "String", "[" and "]". Within a string pattern, "∨" is used to match the pattern that follows with the list zero or more characters.

For example:

- String[x, '2', ∨y] is a pattern that matches "1234" and binds x to '1' and y to "34".
- String['1', '2', ∨$$y] is a pattern that only matches "1234" if y is "34".
- String[∨x, ∨y] is an illegal pattern because it can match ambiguously.

As an example of the use of spread, the following procedure reverses a string:[70]

| Symbols |
|---|
| ≡ ꞉ ∨ |
| ⍰ ¶ § ▮ |

Reverse.[aString:String]:String ≡
  aString �
    String[ ] ꞉ String[ ],
    String[first, ∨rest] ꞉ String[rest, first] ⍰▮

The formal syntax of string expressions is in the following end note: **71.**

**General Messaging, *i.e.*, . and ⊡**

The syntax for general messaging is to use an expression for the recipient followed by "." and an expression for the message.

For example, if anExpression is of type Expression◁Integer▷ then,
  anExpression.eval[anEnvironment]▮
is equivalent to the following:
  Let aMessage[i] ← eval⊡Expression◁Integer▷[anEnvironment],
    anExpression.aMessage▮

The formal syntax of general messaging is in the following end note: **72.**

---

[i] aMessage:Message◁Expression◁Integer▷▷

**Language extension,** *i.e.*, ⦅ ⦆

The following is an illustration of language extension that illustrates postponed execution:[i]

⦅“**Postpone**” anExpression:Expression ◁aType▷⦆:Postpone◁aType▷ ≡
  **Actor implements** Expression◁Future◁aType▷▷ **using**
    eval[anEnvironment]→
      **Future Actor implements** aType **using**
        aMessage[ii]→    // aMessage received
          **Let** postponed:aType ←
                anExpression.eval[anEnvironment],
            postponed.aMessage
                // return result of sending aMessage to postponed
              **become** postponed§▮
                // become the Actor postponed for
                  // the next message received[iii]

The formal syntax of language extension is in the following end note: **73**.

---

[i] A **Postpone** expression does not begin execution of $Expression_1$ until a request is received. Illustration:
IntegersBeginningWith.[n:Integer]:◁Future◁◁List◁Integer▷▷ ≡
    [n, ∨**Postpone** IntegersBeginningWith.[n+1]]▮
Note: A **Postpone** expression can limit performance by preventing concurrency

[ii] aMessage:Message◁aType▷

[iii] this is allowed because postponed is of type aType

**Atomic Operations,** *i.e.* **Atomic compare update then else**
For example, the following example implements a lockable that spins to lock:[74]

| Symbols |
|---|
| ≡ → |
| ⍰ ¶ § ▮ |

```
Actor SpinLock[ ] unserialized
     locked ≔ False,               // initially unlocked
     implements Lockable[i] using
       lock[ ]→ Attempt.[ ] ≜        // perform the loop Attempt as follows
                 Atomic locked compare False update True
                   // attempt to atomically update locked from False to True
                   then Precondition locked= True,
                                              // locked must have contents True
                                   Void,     // if updated return Void
                   else Attempt.[ ]¶        // if not updated perform Attempt
       unLock[ ]→
         Precondition locked = True,      // locked must have contents True
            Void afterward locked ≔ False §▮    // reset locked to False
```

The formal syntax of atomic operations is in the following end note: **75.**

**Enumerations,** *i.e.,* **Enumeration { } using Qualifiers,** *i.e.,* ⊡
An enumeration provides symbolic names for alternatives. For example,

```
Enumeration DayName {Monday, Tuesday, Wednesday, Thursday,
                                  Friday, Saturday, Sunday}▮
```

From the above definition, an enumerated day is available using a qualifier, *e.g.,* Monday⊡DayName. Qualifiers provide structure for namespaces.

The formal syntax of qualifiers is in the following end note: **76**.

---

[i] Interface Lockable {lock[ ]↦ Void,
unLock[ ]↦ Void}▮

The procedure below computes the name of following day of the week given the name of any day of the week:

```
UsingNamespace DayName▮
FollowingDay.[aDay:DayName]:DayName ≡
  aDay � Monday ⦂ Tuesday,
          Tuesday ⦂ Wednesday,
          Wednesday ⦂ Thursday,
          Thursday ⦂ Friday,
          Friday ⦂  Saturday,
          Saturday ⦂ Sunday,
          Sunday ⦂ Monday ⍰▮
```

The formal syntax of enumerations is in the following end note: **77.**

**Native types, e.g., JavaScript, JSON, Java, and XML**

**Object** can be used to create JavaScript Objects. Also, **Function** can be used to bind the reserved identifier **This**. For example, consider the following ActorScript for creating a JavaScript object aRectangle (with length 3 and width 4) and then computing its area 12:

```
Let {aRectangle[i] ← Object {"length": 3, "width": 4]},
     aFunction ← Function [ ]→ This⟦"length"⟧ * This⟦"width"⟧},
  Do aRectangle⟦"area"⟧ ≔ aFunction●
    aRectangle⟦"area"⟧.[ ]▮
```

The setTimeout JavaScript object can be invoked with a callback as follows that logs the string "later" after a time out of 1000:

```
setTimeout◻JavaScript.[1000,
                        Function [ ]→
                           console◻JavaScript.["log"].["later"]]▮
```

**JSON** is a restricted version of **Object** that allows only Booleans, numbers, strings in objects and arrays.[ii]

Native types can also be used from Java. For example (s:String◻Java).substring[3, 5][iii] is the substring of s from the 3rd to the 5th characters inclusive.

---

[i] aRectangle is of type Object◻JavaScript

[ii] *i.e.* the following JavaScript types are not included in JSON: Date, Error, Regular Expression, and Function.

[iii] substring is a method of the String class in Java

Java types can be imported using **Import**, *e.g.*:

**Namespace** mynamespace;
**Import** java.math.BigInteger;
**Import** java.lang.Number;

After the above, **BigInteger.new**["123"]**.instanceof[Number]▮** is equivalent to **True▮**:

The following notation is used for XML:[78]

**XML** <“PersonName”> <“First”>“Ole-Johan” </“First”>
<“Last”> **“**Dahl”</“Last”> </“PersonName”>

and could print as:

<PersonName> <First> Ole-Johan </First>
<Last> Dahl </Last> </PersonName>

XML Attributes are allowed so that the expression

**XML** <“Country” “capital”=“Paris”> “France” </“Country”>

and could print as:

<Country capital=**“**Paris**”**> France </Country>

XML construction can be performed in the following ways using the append operator:

- **XML** <“doc”> 1, 2, ⩛[3] </“doc”>]▮ is equivalent to **XML** <“doc”>1, 2, 3</“doc”>▮
- **XML** <“doc”>1, 2, ⩛[3], ⩛[4] </“doc”>]▮ is equivalent to **XML** <“doc”> 1, 2, 3, 4 </“doc”>▮

**One-way messaging, *e.g.*, ⊖, ↞, and ↠**

One-way messaging is often used in hardware implementations.

Each one-way named-message send consists of an expression followed by "↞", a message name, and arguments delimited by "[" and "]".

The following is a interface for a customer that is used in request/response message passing for return type aType:[79]

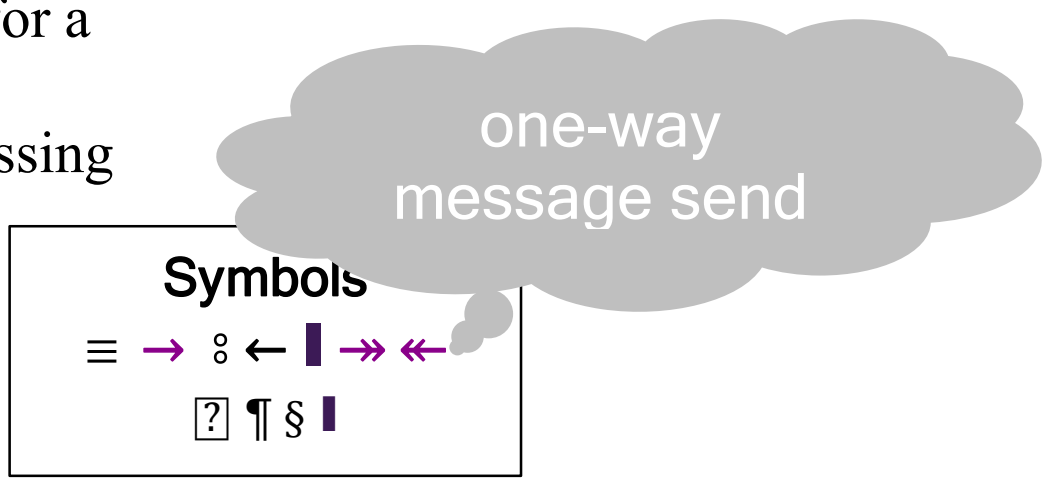

Interface Customer◁aType▷
  {return[aType] ↦ ⊖,
  throw[anException] ↦ ⊖}

For example, if aCustomer is of type Customer◁Integer▷, then 3 can be returned to aCustomer using aCustomer↞return[3].

The formal syntactic definition of one-way named-message sending is in the end note: **80**

Each one-way message handler implementation consists of a named-message declaration pattern followed by "↠" and a body for the response which must ultimately be "⊖" which denotes no response.

The formal syntactic definition of one-way named-message implementation is in the following end note: **81**

The following is an implementation of an arithmetic logic unit that implements **jumpGreater** and **addJumpPositive** one-way messages:

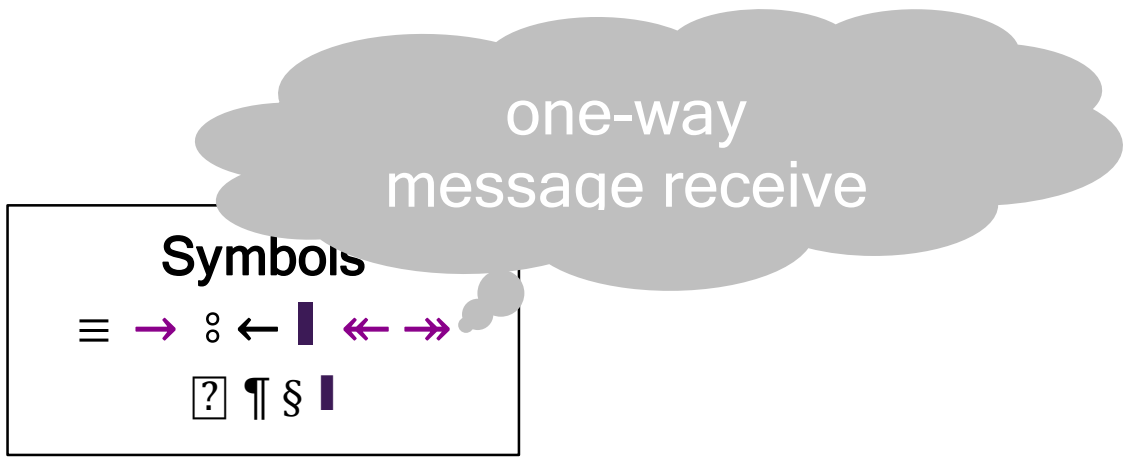

```
Actor ArithmeticLogicUnit◁aType▷[ ]
 implements ALU◁aType▷[i] using
  jumpGreater[x:aType, y:aType,
                 firstGreaterAddress:Address, elseAddress:Address]↠
       InstructionUnit↞Execute[(x>y) �
           True ⸽ firstGreaterAddress,
           else ⸽ elseAddress ▯]¶
  addJumpPositive[x:aType, y:aType, sumLocation:Location◁aType▷,
                      positiveAddress:Address, elseAddress:Address]↠
       Let z ← (x+y),
        sumLocation �
          aVariableLocation:VariableLocation◁aType▷[ii] ⸽
            Do aVariableLocation.store[z] ●
                        // continue after acknowledgement of store
              (z >0) � True ⸽ InstructionUnit↞execute[positiveAddress],
                         False ⸽ InstructionUnit↞execute[elseAddress] ▯,
          aTemporaryLocation:TemporaryLocation◁aType▷[iii] ⸽
            Do aTemporaryLocation↞write[z],
                      // continue concurrently with processing write
              (z >0) � True ⸽ InstructionUnit↞execute[positiveAddress]
                         False ⸽ InstructionUnit↞execute[elseAddress] ▯ ▯§▮
```

**Arrays**

Arrays are lists of locations that can be updated using **swap** messages.

They are included to provide backward compatibility and to support certain kinds of low level optimizations. An array element can be referenced using the array followed by array indices enclosed by "⟦" and "⟧".

---

[i] **Interface** ALU◁aType▷ {jumpGreater [aType, ] ↦ ⊖,
addJumpPositive [anException] ↦ ⊖}

[ii] VariableLocation◁aType▷ **has** store[aType]↦ **Void**

[iii] TemporaryLocation◁aType▷ **has** write[aType] ↦ ⊖

In the in-place implementation of QuickSort[82] (below), left is the index of the leftmost element of the subarray, right is the index of the rightmost element of the subarray (inclusive), and the number of elements in the subarray is right-(left+1).

```
QuickSort.[anArray:Array◁Number▷, left:Integer, right:Integer]:Void ≡
  Precondition anArray.lower[ ]≦left≦right≦anArray.upper[ ],
   (left<right) � True ⦂                          // If the array has 2 or more items
                      Let pivotIndex ←
                          Partition.[anArray, left, right, left+(right-left)/2]●
                        Precondition  left≦pivotIndex≦right,
                          Do ⦶QuickSort.[anArray, left, pivotIndex-1],
                              // Recursively sort elements smaller than the pivot
                            ⦶QuickSort.[anArray, pivotIndex+1, right]
                                     // Concurrently recursively sort elements at
                                               //   least as big as pivot
                    False ⦂ Void ⍰▮
```

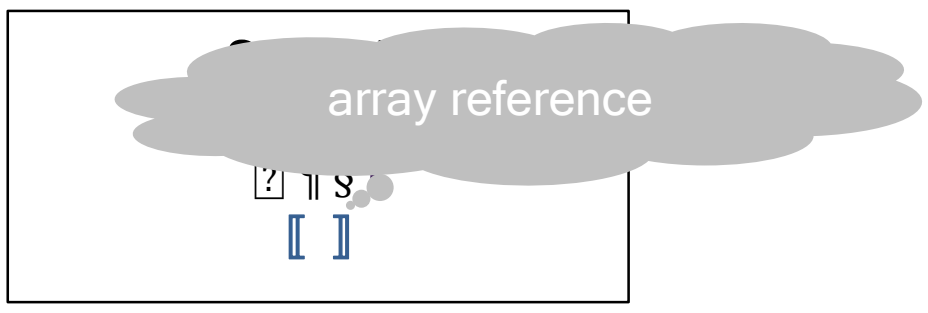

```
Partition.[anArray:Array◁Number▷, left:Integer, right:Integer,
          pivotIndex:Integer]:Integer ≡
  Precondition anArray.lower[ ]≦left≦pivotIndex≦right≦anArray.upper[ ],
    Let pivot← anArray⟦pivotIndex⟧●
      Do anArray.swap[pivotIndex, right]●[83]
        Let finalStoreIndex ←
              Move.[iterationIndex:Integer ← left,
                      storeIndex:Integer ←left]:Integer ≜
                Precondition  left≦storeIndex≦iterationIndex≦right,
                 iterationIndex<right �
                    True ⦂
                       anArray⟦iterationIndex⟧≦pivotValue �
                         True ⦂
                           Do anArray.swap[iterationIndex, storeIndex]●
                              Move.[iterationIndex+1, storeIndex+1],
                         False ⦂ Move.[iterationIndex+1, storeIndex] ⍰,
                    False ⦂ storeIndex ⍰●
          Do anArray.swap[finalStoreIndex, right]●
                          // Move Actor stored at right to its final place
            finalStoreIndex▮
```

For example, consider the following example:

```
        Let anArray ← Array.[3, 2, 1],
          Do QuickSort.[anArray, 0, 1]●
            anArray▮
```

The above returns Array[1, 2, 3]▮.

**Extending Implementations, *i.e.*, extends**

Consider the following implementation of Account:

```
Actor SimpleAccount◁aCurrency⊑Currency▷[aBalance:aCurrency]
   myBalance ≔ aBalance,
   implements Account◁aCurrency▷[84]  using
     getBalance[ ]:aCurrency → myBalance¶
     deposit[anAmount] →
       Void afterward  myBalance ≔ myBalance+anAmount¶
     withdraw[anAmount] →
       (anAmount > myBalance) �
          True ⦂ Throw  OverdrawnException[ ],
          False ⦂
            Void afterward myBalance ≔ myBalance–anAmount ⍰§▮
```

A *facet*[85] is an address of an Actor whose type is of a subsidiary interface of the Actor which is expressed using "⦂" followed by the name of the interface.

An implementation can be extended using "**extends**" followed by a constructor. For example, the implementation Account above can be extended as follows:

```
Actor FeeAccount◁aCurrency⊑Currency▷[initialBalance:aCurrency,
                                      fee:aCurrency]
                extends SimpleAccount◁aCurrency▷[initialBalance]
   partially reimplements Account◁aCurrency▷ using
     withdraw[anAmount:aCurrency]:Void override→
       Do ⦂Account▪withdraw[anAmount]⬣
          // Use Account for delegated withdraw of
          //   anAmount from this account.
               // Note that  myBalance changes.
        ⦂Account▪withdraw[fee]§▮
          // return delegated withdraw of fee from this account
```

Please note the following:

- FeeAccount◁Euro▷.[€3]:Account▮ is equivalent to **False**▮ because a FeeAccount is not of exact type Account
- FeeAccount◁Euro▷.[€3]::Account▮ is equivalent to **True**▮ because a FeeAccount is of extended type of Account

Also, the implementation Account can be extended as follows to provide the ability to revoke abilities to change an account. For example, SimpleAccountMonitor below implements both the Account◁aCurrency▷ and AccountRevoker interfaces as an extension of the implementation SimpleAccount:

```
Actor SimpleMonitor◁aCurrency⊑Currency▷ [initialBalance:aCurrency]
 extends SimpleAccount◁aCurrency▷[initialBalance]
   withdrawIsRevoked ≔ False,
   depositIsRevoked ≔ False,
   implements AccountMonitor◁aCurrency▷[86] using
     getRevoker[ ]→ ⍠AccountRevoker¶
     getAccount[ ] → ⍠Account◁aCurrency▷¶
     withdrawFee[anAmount] →
            Void afterward myBalance ≔ myBalance−anAmount§
                   // withdraw fee even if balance goes negative
   also partially reimplements Account◁aCurrency▷ using
     withdraw[anAmount:aCurrency] →
       withdrawIsRevoked �
         True ⦂ Throw Revoked[ ],
         False ⦂ ⍠Account◁aCurrency▷▪withdraw[anAmount] ⍰¶
     deposit[anAmount:aCurrency] →
       depositIsRevoked �
         True ⦂ Throw Revoked[ ],
         False ⦂ ⍠Account◁aCurrency▷▪deposit[anAmount] ⍰§
   also implements AccountRevoker[87] using
     revokeDeposit[ ] →
       Void afterward depositIsRevoked ≔ True¶
     revokeWithdraw[ ] →
       Void afterward withdrawIsRevoked ≔ True§▮
```

For example, the following expression returns €5:

```
 Let aMonitor ← SimpleAccountMonitor◁Euro▷.[€3],
  Let {anAccount ← aMonitor▪getAccount[ ],
       aRevoker ← aMonitor▪getRevoker[ ]},
    Do [anAccount▪deposit[€2]⚫      // the balance in anAccount is €5
        aRevoker▪revokeDeposit[ ]⚫
                      // depostIsRevoked in aMonitor is True
        Try anAccount▪deposit[€5]              // try another deposit
          catch� _ ⦂ Void] ⍰]⚫           // ignore the thrown exception
      anAccount▪getBalance[ ]▮  // the balance in anAccount remains €5
```

## Appendix 2: Meta-circular definition of ActorScript

It might seem that a meta-circular definition is a strange way to define a programming language. However, as shown in the references, concurrent programming languages are not reducible to logic. Consequently, an augmented meta-circular definition may be one of the best alternatives available.

### The message eval

John McCarthy is justly famous for Lisp. One of the more remarkable aspects of Lisp was the definition of its interpreter (called Eval) in Lisp itself. The exact meaning of Eval defined in terms of itself has been somewhat mysterious since, on the face of it, the definition is circular.[88]

The basic idea is to send an expression an **eval** message with an environment to instead of the Lisp approach of send the procedure Eval the expression and environment as arguments.

Each **eval** message has an environment with the bindings of program identifiers:

```
    Interface Expression◁aType▷
       extends Construct◁aType▷
                eval[Environment]↦ aType▮
```

The tokens ⦅ and ⦆ are used to delimit program syntax.

```
⦅anIdentifier:Identifier◁aType▷⦆:Expression◁aType▷ ≡
  eval[anEnvironment]→ anEnvironment.lookup[anIdentifier]▮
```

### Querying an Actor using isOfExactly◻Query and isOfAnExtension◻Query

There is a special distinguished message named **isOfExactly◻Query** that can be used to query whether an Actor implements a type by sending it a hash for the type.

```
⦅anExpression:Expression◁aType▷ ":" aTypeExpression:Type⦆
                                         :Expression◁Boolean▷ ≡
  eval[anEnvironment]→
     anExpression.eval[anEnvironment]
                 .isOfExactly◻Query[aTypeExpression
                                      .eval[anEnvironment]]▮
              // anExpression implements aTypeExpression
```

In addition, there is a special distinguished message named isOfAnExtension⊡Query that can be used to query whether an Actor inherits from a type by sending it a hash for the type.

```
⦅anExpression:Expression◁aType▷ "::" aTypeExpression:Type⦆
                                        :Expression◁Boolean▷ ≡
   eval[anEnvironment]→
      anExpression▪eval[anEnvironment]
                  ▪isOfAnExtension⊡Query
                     [aTypeExpression▪eval[anEnvironment]]▮
              // anExpression implements aTypeExpression
```

**The interface Type implements isExtension**

The interface Type has message isExtension:

Interface Type {isExtension[aType:Type]↦ Boolean}▮

```
⦅anotherType:Type◁anotherType▷ "⊒" aType:Type◁aType▷⦆
                                        :Expression◁Boolean▷ ≡
  eval[anEnvironment]→
     (anotherType▪eval[anEnvironment])
           ▪isExtension[aType▪eval[anEnvironment]]▮
```

**Future, ↓, and ⓕ**

The interface Future is used for futures:

Interface Future◁aType▷ {resolve[ ]↦ aType}▮

```
⦅"Future" anExpression:Expression◁aType▷⦆
                                   :Expression◁Future◁aType▷▷ ≡
  Actor implements Expression◁Future◁aType▷▷ using
    eval[anEnvironment]→
      Let aFuture:Future◁aType▷ ←
            Future Try anExpression.eval[anEnvironment]
                     catch�
                       anException ∷
                          Actor
                            implements Future◁aType▷
                              {resolve[ ]→Throw anException}⍰
        Actor implements Future◁aType▷using
              resolve[ ]→ ↓aFuture §▮
```

```
⦅"↓" anExpression:Expression◁Future◁aType▷▷⦆
                                   :Expression◁aType▷ ≡
  Actor implements Expression◁aType▷ using
    eval[anEnvironment]→
        anExpression.eval[anEnvironment].resolve[ ] §▮
```

```
⦅"ⓕ" anExpression:Expression◁aType▷⦆
                                   :Expression◁aType▷ ≡
  Actor implements Expression◁aType▷ using
    eval[anEnvironment]→
        ↓Future anExpression.eval[anEnvironment] §▮
```

**The message match**

Patterns are analogous to expressions, except that they take receive match messages:

Interface Pattern◁aType▷
  {match[aType, Environment]↦ Nullable◁Environment▷,
   mustMatch[aType, Environment]↦ Environment}▮

```
⦅anIdentifier:Identifier ◁aType▷⦆:Pattern◁aType▷  ≡
     match[anActor:aType, anEnvironment]→
          anEnvironment▪bind[anIdentifier,  to ⌸ anActor]▮
```

```
⦅“_”⦆:UniversalPattern◁aType▷ ≡
   match[anActor:aType, anEnvironment]→ anEnvironment▮
```

```
⦅aPattern:Pattern◁aType▷ “:” aTypeExpression:Type⦆
                                             :TypedPattern◁aType▷  ≡
    match[anActor, anEnvironment]→
      anActor▪[aTypeExpression▪eval[anEnvironment]
                  transitive ⌸ False] �
        True ⦂            // anActor directly implements aTypeExpression
          aPattern▪match[anActor, anEnvironment],
        False  ⦂ Null Environment ⍰▮
```

```
⦅“$$” anExpression:Expression◁expressionType▷⦆
                                            :ValuePattern◁aType▷ ≡
   match[anActor:aType, anEnvironment]→
     (anActor = anExpression▪eval[anEnvironment]) �
        True ⦂ anEnvironment,
        False ⦂ Null Environment ⍰▮
```

**Message sending, *e.g.*, ▪**

```
⦅procedure:Expression◁argumentsType↦returnType▷
     "▪" "[" arguments:Arguments◁argumentsType▷ "]"
        ":" returnType⦆
               :ProcedureSend◁argumentsType, returnType▷ ≡
  eval[anEnvironment]→
    (procedure▪eval[anEnvironment])
        ▪[∀(expressions▪eval[anEnvironment])]§▮
```

```
⦅recipient:Expression◁recipient▷ "▪"
    name:MessageName"[" ∀arguments
                                   :Arguments◁argumentsType▷ "]"⦆
                            :NamedMessageSend◁expressionType▷ ≡
  eval[anEnvironment]→
    Let aRecipient ← recipient▪eval[anEnvironment],
     aRecipient
       ▪Message[QualifiedName[name recipientType],
                 [∀arguments▪eval[anEnvironment]]]§▮
```

```
⦅recipient:Expression◁recipientType▷
     "▪" aMessage:Message◁Message◁recipientType▷▷⦆
                          :UnnamedMessageSend◁expressionType▷ ≡
  eval[anEnvironment]→
    (recipient▪eval[anEnvironment])
       ▪(aMessage▪eval[anEnvironment])§▮
```

**List Expressions and Patterns**

```
(|"[" firstExpression:Expression◁type1▷
     "," secondExpression:Expression◁type2▷"]"|)
                                   :ListExpression◁expressionType▷ ≡
  eval[anEnvironment]→
    Let {first ← firstExpression.eval[anEnvironment],
         second ← secondExpression.eval[anEnvironment]}
      [first, second]§▮
```

```
(|"[" firstExpression:Expression◁type1▷
   "," "⩛" restExpression:Expression◁type2▷ "]"|)
                                   :ListExpression◁expressionType▷ ≡
 eval[anEnvironment]→
    Let {first ← firstExpression.eval[anEnvironment],
         rest ← restExpression.eval[anEnvironment]}
      [first, ⩛rest]§▮
```

```
(|"[" firstPattern:Pattern◁type1▷
                         "," "⩛" restPattern:Pattern◁type1▷ "]"|)
                                         :ListPattern◁aType▷ ≡
 match[anActor:aType, anEnvironment]→
    anActor � [first:type1, ⩛rest:type2] ⦂
       firstPattern.match[first, anEnvironment] �
          Null Environment ⦂ Null Environment,
           aNewEnvironment:Environment ⦂
               restPattern.match[restValue, aNewEnvironment] ⍰,
        else ⦂ Null Environment ⍰§▮
```

### Exceptions

```
⦅"Try" anExpression:Expression◁aType▷
  "catch�"  exceptions:ExpressionCases◁aType▷ "⍰"⦆
                                          :TryExpression◁aType▷ ≡
 eval[anEnvironment]→
     Try anExpression▪eval[anEnvironment] catch�
       anException:Exception ⦂ CasesEval▪[anException,
                                          exceptions,
                                          anEnvironment] ⍰§▮
```

```
⦅"Try" anExpression:Expression◁aType▷
        "cleanup" aCleanup:Expression◁aType▷⦆
                                         :TryExpression◁aType▷ ≡
   eval[anEnvironment]→
      Try anExpression▪eval[anEnvironment] catch�
       _:Exception ⦂ Do aCleanup▪eval[anEnvironment];
                     Rethrow⍰§▮
```

### Continuations using perform

A continuations is a generalization of expression for executing in cheese, which receives perform messages:

```
  Interface Continuation◁aType▷ extends Construct◁aType▷
        perform [Environment, CheeseQ]↦ aType▮
```

```
Execute▪[aConstruct:Construct◁aType▷,
         anEnvironment:Environment,
         aCheeseQ:CheeseQ] ≡
 aConstruct � aContinuation:Continuation◁aType▷ ⦂
                 aContinuaton▪perform[anEnvironment,
                                      aCheeseQ],
               else anExpression:Expression◁aType▷ ⦂
                 anExpression▪eval[anEnvironment] ⍰▮
```

**Atomic compare and update**

```
⟬"Atomic" location:Expression◁Location◁anotherType▷▷,
              "compare"  comparison:Expression◁anotherType▷
              "update" update:Expression◁anotherType▷
              "then"  compareIdentical:Continuation◁aType▷
               "else"  compareNotIdentical:Continuation◁aType▷⟭
                                                          :Atomic◁aType▷ ≡
   perform[anEnvironment, aCheeseQ]→
     (location.eval[anEnvironment])
       .compareAndConditionallyUpdate[comparison.eval[anEnvironment],
                                       update.eval[anEnvironment]] �
         True ⦂ compareIdentifical.perform[anEnvironment, aCheeseQ],
         False ⦂
           compareNotIdentifical.perform[anEnvironment, aCheeseQ] ⍰▮

Actor SimpleLocation◁anotherType▷[initialContents]
   contents ≔ initialContents,
   implements Location◁anotherType▷ using
     compareAndConditionallyUpdate[comparison, update]→
       (contents = comparison) �
           True ⦂ True afterward contents ≔ update,
           False ⦂ False ⍰§▮
```

**Cases**

```
⦅anExpression:Expression◁anotherType▷ "�"
                    cases:ExpressionCases◁aType▷ "⍰"⦆
                                        :CasesExpression◁aType▷ ≡
   eval[anEnvironment]→
      CasesEval▪[anExpression▪eval[anEnvironment],
                 cases,
                 anEnvironment]§▮

CasesEval▪[anActor:anotherType,
           cases:List◁ExpressionCase◁aType▷▷,
           anEnvironment:Environment] ≡
 cases �
   [ ] ⦂ Throw NoApplicableCase[ ],
   [first, ⩛rest] ⦂
     first � ⦅aPattern:Pattern◁anotherType▷ "⦂"
                anExpression:Expression◁aType▷⦆
                                    :ExpressionCase◁aType▷ ⦂
             aPattern▪match[anActor, anEnvironment] �
               Null Environment ⦂
                 CasesEval▪[anActor, rest, anEnvironment],
               newEnvironment:Environment ⦂
                 anExpression▪eval[newEnvironment] ⍰,
           ⦅"else" elsePattern:Pattern◁anotherType▷"⦂"
                elseExpression:Expression◁aType▷⦆
                                    :ExpressionElseCase◁aType▷ ⦂
             elsePattern▪match[anActor, anEnvironment] �
               Null Environment ⦂
                   Throw ElsePatternMustMatch[ ],
               newEnvironment:Environment ⦂
                   elseExpression▪eval[newEnvironment] ⍰,
           ⦅"else" "⦂"
               elseExpression:Expression◁aType▷⦆
                                    :ExpressionElseCase◁aType▷ ⦂
             elseExpression▪eval[anEnvironment],
           else ⦂ Throw NoApplicableCase[ ] ⍰⍰▮
```

```
⦅anExpression:Expression◁anotherType▷ "�"
                    cases:ContinuationCases◁aType▷ "⍰"⦆
                                        :CasesContinuation◁aType▷ ≡
   perform[anEnvironment, aCheeseQ]→
      CasesPerform▪[anExpression▪eval[anEnvironment], cases,
                   anEnvironment, aCheeseQ]§▮
CasesPerform▪ [anActor:anotherType,
                    cases:List◁ContinuationCase◁aType▷▷,
                    anEnvironment:Environment,
                    aCheeseQ:CheeseQ] ≡
 cases �
  [ ] ⦂ Throw NoApplicableCase[ ],
  [first, ⩛rest] ⦂
   first �  ⦅aPattern:Pattern◁anotherType▷"⦂"
                    aContinuation:Continuation◁aType▷⦆
                                        :ContinuationCase◁aType▷ ⦂
                 aPattern▪match[anActor, anEnvironment] �
                    Null Environment ⦂
                      CasesPerform▪[anActor,
                                     rest,
                                      anEnvironment,
                                      aCheeseQ],
                    newEnvironment:Environment ⦂
                      aContinuation▪perform[newEnvironment,
                                           aCheeseQ] ⍰,
               ⦅"else"
                    elsePattern:Pattern
                     ◁anotherType▷ "⦂"
                        elseContinuation:Continuation◁aType▷⦆
                                       :ContinuationElseCase◁aType▷ ⦂
                      elsePattern▪match[anActor, anEnvironment] �
                        Null Environment ⦂
                           Throw ElsePatternMustMatch[ ],
                         newEnvironment:Environment ⦂
                             elseContinuation▪eval[newEnvironment] ⍰,
               ⦅"else" "⦂"
                    elseContinuation:Continuation◁aType▷⦆
                                       :ContinuationElseCase◁aType▷ ⦂
                    elseContinuation▪perform[anEnvironment, aCheeseQ],
               else ⦂ Throw NoApplicableCase[ ] ⍰⍰▮
```

### Holes in the cheese

```
⦅anExpression:Expression◁aType▷
   "afterward" "{"someAssignments:Assignments"}"⦆
                                          :Continuation◁aType▷ ≡
  perform[anEnvironment, aCheeseQ]→
    Let anActor ← anExpression.eval[anEnvironment]●
      Do [someAssignments.carryOut[anEnvironment,
                                   aCheeseQ]●
         aCheeseQ.leave[ ]]●
        anActor§▮
```

```
⦅aVariable:Variable◁aType▷
          ":=" anExpression:Expression◁aType▷⦆:Assignment ≡
 carryOut[anEnvironment]→
   anEnvironment.assign[aVariable,
                        to ⌸ anEpression.eval[anEnvironment]]§▮
```

```
⦅"Hole" anExpression:Expression◁aType▷⦆:Hole◁aType▷ ≡
  perform[anEnvironment, aCheeseQ]→
    Do aCheeseQ.leave[ ]●
      anExpression.eval[anEnvironment.freeze[ ]]§▮
```

```
⦅"Hole" anExpression:Expression◁aType▷
   "after"
     aPreparation:Preparation⦆:Continuation◁aType▷ ≡
  perform[anEnvironment, aCheeseQ]→
    Let frozenEnvironment ← anEnvironment.freeze[ ]●
      // create frozen environment so that
         // preparation does not affect evaluating anExpression
      Do [aPreparation.carryOut[anEnvironment, aCheeseQ]●
          aCheeseQ.leave[ ]]●
        anExpression.eval[frozenEnvironment]§▮
```

```
⁅"Hole" anExpression:Expression◁anotherType▷
   "afterward" anAfterward:AfterwardContinuation◁aType▷ "?"⁆
                                          :Continuation◁aType▷ ≡
  perform[anEnvironment, aCheeseQ]→
    Let frozenEnvironment ← anEnvironment.freeze[ ]●
      Do aCheeseQ.leave[ ]●
        Try Let anActor ← anExpression.eval[frozenEnvironment]●
              Do [aCheeseQ.enter[ ]●
                  anAfterward.perform[anEnvironment,
                                      aCheeseQ]]●
                anActor afterward aCheeseQ.leave[ ]
            catch� anException:ApplicationException ⦂
              Do [aCheeseQ.enter[ ]●
                  anAfterward.perform[anEnvironment,
                                      aCheeseQ]]●
                 throw anException afterward aCheeseQ.leave[ ]§▮
```

```
⁅"Hole" anExpression:Expression◁anotherType▷
   "returned�" returnedCases:ContinuationCases◁aType▷ "?"
    "threw�" threwCases:ContinuationCases◁aType▷ "?" ⁆
                                          :Continuation◁aType▷ ≡
  perform[anEnvironment, aCheeseQ]→
    Let frozenEnvironment ← anEnvironment.freeze[ ]●
      Do aCheeseQ.leave[ ]●
        Try Let anActor ← anExpression.eval[frozenEnvironment]●
              Do aCheeseQ.enter[ ]●
                CasesPerform.[anActor,
                              returnedCases,
                              anEnvironment,
                              aCheeseQ]
            catch� anException:ApplicationException ⦂
              Do aCheeseQ.enter[ ]●
                CasesPerform.[anException,
                              threwCases,
                              anEnvironment,
                              aCheeseQ]§▮
```

```
⟦"Enqueue" anExpression:QueueExpression "●"⟧
                                        :Enqueue◁aType▷ ≡
  perform[anEnvironment, aCheeseQ]→
    Let anInternalQ ← anExpression.eval[anEnvironment],
      anInternalQ.enqueueAndLeave[ ] ⍰§▮
```

```
⟦"Enqueue" anExpression:QueueExpression "●"
    aContinuation:Continuation◁aType▷⟧:Enqueue◁aType▷ ≡
  perform[anEnvironment, aCheeseQ]→
    Let anInternalQ ← anExpression.eval[anEnvironment],
     Do anInternalQ.enqueueAndLeave[ ]●
      aContinuation.perform[anEnvironment, aCheeseQ] ⍰§▮
```

**A Simple Implementation of Actor**

The implementation below does not implement queues, holes, and relaying.

```
⟦"Actor" declarations:ActorDeclarations
    "implements" Identifier◁aType▷
    "using" handlers:Handlers◁anInterface▷ "§"⟧:Actor◁aType▷ ≡
  Actor implements Expression◁anInterface▷ using
   eval[anEnvironment]→
     Initialized.[anInterface.eval[anEnvironment],
                                   handlers,
                                   declarations
                                     .initialize[anEnvironment],
                                         SimpleCheeseQ.[ ]]§▮
```

There are special distinguished messages named isOfExactly and isOfAnExtension that can be used to query an Actor whether it exactly implements a type or implements and an extension of a type.

```
Initialized.[anInterface:aType,
             handlers:List◁Handler◁aType▷▷,
             anEnvironment:Environment,
             cheeseQ:CheeseQ] ≡
  Actor implements anInterface using
    receivedMessage:Message◁aType▷ →
                                   // receivedMessage received for anInterface
      receivedMessage �
         isOfExactly□Query[aType] ⦂
            aType=anInterface,
                           // test if this Actor implements anotherType
         isOfAnExtension□Query[aType] ⦂
            aType=anInterface �
               True ⦂ True,
               False ⦂ aType.isExtension[anInterface]⍰⍰ ,
         else ⦂
           Do aCheeseQ.enter[ ]●
              Let aReturned ← Try Select.[receivedMessage,
                                          handlers,
                                          anEnvironment,
                                          aCheeseQ]
                                  cleanup aCheeseQ.leave[ ]●
                                    // leave cheese and rethrow exception
                 Do aCheeseQ.leave[ ]●
                    aReturned▮
```

```
Select▪[receivedMessage:Message◁aType▷,
         handlers:List◁Handler◁aType▷▷,
         anEnvironment:Environment,
         aCheeseQ:CheeseQ]:aType ≡
   handlers �
     [ ] ⦂ Throw NotApplicable[ ],
     [⦅aMessageDeclaration:MessageDeclaration◁aType▷ "→"
          body:Continuation◁aType▷⦆
                                          :ContinuationHandler◁aType▷,
      ∨restHandlers] ⦂
         aMessageDeclaration▪match[receivedMessage,
                                    anEnvironment] �
            Null Environment ⦂ Select▪[receivedMessage,
                                        restHandlers,
                                        anEnvironment,
                                        aCheeseQ],
                                 // process next handler
            newEnvironment:Environment ⦂
               Execute▪[body, newEnvironment, aCheeseQ] ?? ▌
                        // execute body with extension of anEnvironment
```

### An implementation of cheese that never holds a lock

The following is an implementation of cheese that does not hold a lock:[89]

```
 Actor SimpleCheeseQ[ ] unserialized
     Invariant aTail=Null Activity ⇨ previousToTail=Null Activity
     aHeadHint ≔ Null Activity,            // aHeadHint:Nullable◁Activity▷90
     aTail ≔ Null Activity,                // aTail:Nullable◁Activity▷91
     implements CheeseQ92 using
       enter[ ] in myActivity →93
          Preconditions {myActivity⟦previous⟧=Null Activity,
                         myActivity⟦nextHint⟧=Null Activity},
            attempt.[ ]:Void ≜
              Do myActivity⟦previous⟧ ≔ aTail●   // set provisional tail of queue
               Atomic aTail compare aTail update myActivity
                 then  // inserted myActivity in the cheese queue with previous
                    myActivity⟦previous⟧=Null Activity �
                           True ⦂ Void,
                           False ⦂ Suspend ⍰,  // current activity is suspended
                 else attempt.[ ]¶
       leave[ ] in myActivity →
                                  // leave message received running myActivity
          Precondition aTail != Null Activity,94
            Let ahead ← ⍠SubCheeseQ▪head[ ]●
             Precondition ahead = myActivity,
               Atomic aTail compare ahead update Null Activity
                 then                 // last activity has left this cheese queue
                   Do aHeadHint ≔ Null  Activity●
                    Void,
                 else                 // another activity is in this cheese queue
                   Do aHeadHint ≔ ahead⟦nextHint⟧●
                    MakeRunnable aHeadHintΔActivity;§
     also implements SubCheeseQ95 using
       head[ ] → Precondition aTail != Null  Activity,
         findHead.[backIterator:Activity ←
                     aHeadHint �
                       Null Activity ⦂ aTailΔActivity],
                       else ⦂ aHeadHintΔActivity ⍰]:Activity ≜
           backIterator⟦previous⟧ �
            Null Activity ⦂
                          // backIterator is at the head of the cheese queue
              Do aHeadHint ≔ Nullable backIterator●
               backIterator,
            previousBackIterator:Activity ⦂
                                  //  backIterator is not the head of this CheeseQ
              Do previousBackIterator⟦nextHint⟧ ≔ Nullable backIterator●
                                      // set nextHint of previous to backIterator
               findHead.[previousBackIterator] ⍰§▮
```

The algorithm used in the implementation of **CheeseQ** above is due to Blaine Garst [private communication] *cf.* [Ladan-Mozes and Shavit 2004].

There is a state diagram for the implementation below:

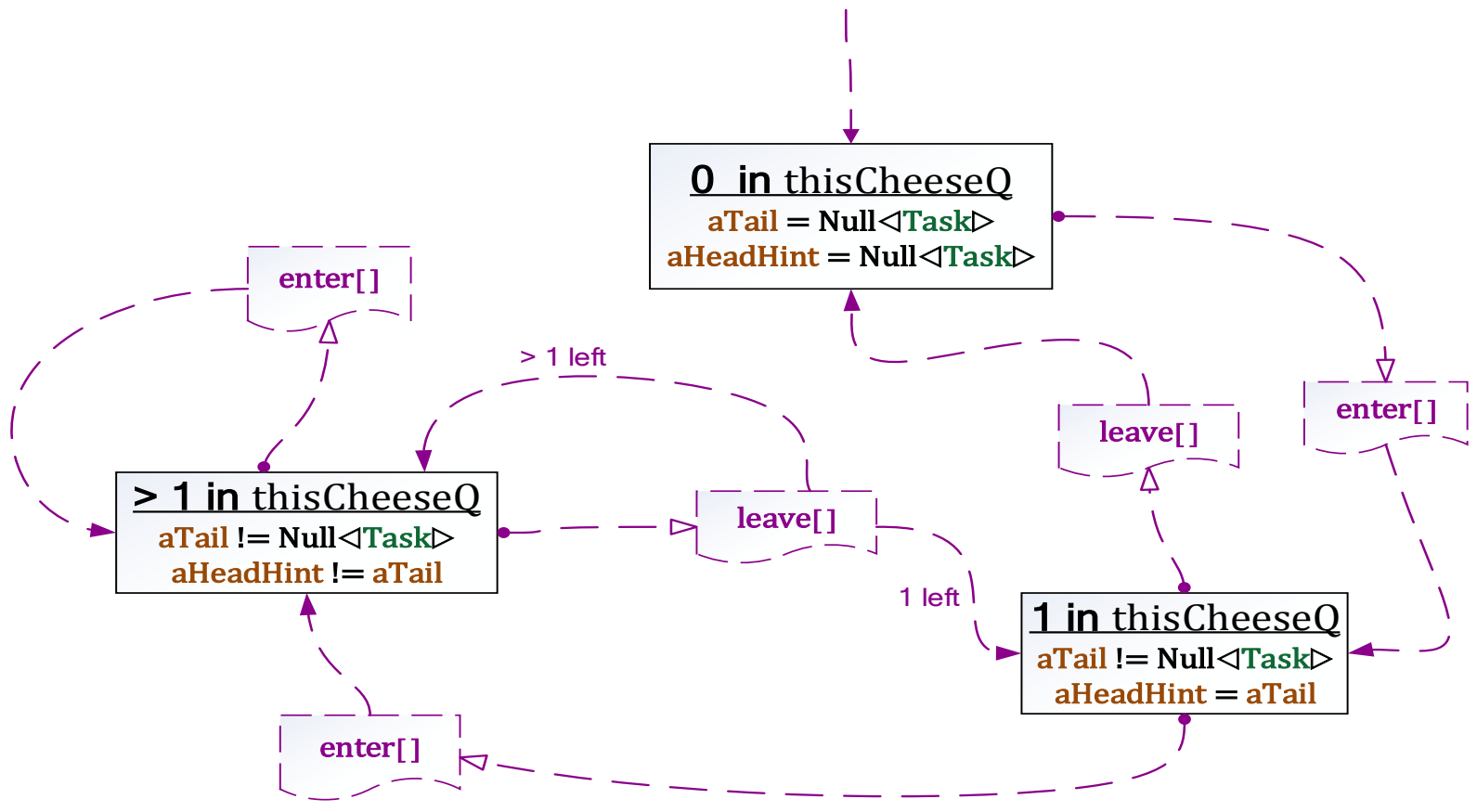

```
Actor SimpleInternalQ [aCheeseQ:CheeseQ] unserialized
  aHead ≔ Null Activity,           // aHead:Nullable◁Activity▷
  aTail ≔ Null Activity,
  implements InternalQ96 using
     enqueueAndLeave[ ]  in myActivity →
                 // enqueueAndLeave message received in myActivity
           Do [⍠subInternalQ▪remove[myActivity]⬣
                aCheeseQ▪leave[ ]]⬣    // myActivity is the head of aCheeseQ
              Suspend
                // myActivity is suspended and when resumed returns Void ¶
     enqueueAndDequeue[anInternalQ] in myActivity →
        Precondition ¬anInternalQ▪isEmpty[ ],
           Do [⍠subInternalQ▪add[myActivity]⬣
                ▪▪dequeue[ ]]⬣
              Suspend¶
     dequeue[ ]  in myActivity → Precondition ¬ ▪▪isEmpty[ ],
              Do aCheeseQ▪leave[ ]⬣          // myActivity is the head of aCheeseQ
                 MakeRunnable ⍠subInternalQ▪remove[ ]¶
                                         // make runnable the removed activity
     isEmpty[ ] → aTail = Null Activity§
  also implements subInternalQ97 using
      add[anActivity] →
        Precondition  anActivity⟦rest⟧ = Null Activity;
              // anActivity must not be in another internal cheese queue
           aTail �
              Null Activity ⦂
                 Void afterward {aHead ≔ Nullable anActivity,
                                 aTail ≔ Nullable anActivity},
              theTail:Activity ⦂ Void afterward theTail⟦rest⟧ ≔ anActivity ⍰¶
      remove[ ]:Activity →  Precondition ¬ ▪▪isEmpty[ ];
        Let theFirst ← aHeadΔActivity⬣
           aTail=aHead �
              True ⦂ theFirst afterward {aHead ≔ Null Activity,
                                         aTail ≔ Null Activity},
               False ⦂ theFirst afterward aHead ≔ (aHeadΔActivity)⟦rest⟧ ⍰§▮
```

## Appendix 3. Inconsistency Robust Logic Programs

Logic Programs[98] can logically infer computational steps.

### Forward Chaining

Forward chaining is performed using ⊢

⟦"⊢"$_{Theory}$ *PropositionExpression*⟧
Assert *PropositionExpression* for *Theory*.

⟦**"When"** "⊢"$_{Theory}$ aProposition:*Pattern* "→" *Expression*⟧
When aProposition holds for *Theory*, evaluate *Expression*.

Illustration of forward chaining:

⊢$_t$ Human[Socrates]▮

**When** ⊢$_t$ Human[$x$] → ⊢$_t$ Mortal[$x$]▮

will result in asserting Mortal[Socrates] for theory t

### Backward Chaining

Backward chaining is performed using ⊩

⟦"⊩"$_{Theory}$ aGoal:*Pattern* "→" *Expression*⟧
Set aGoal for *Theory* and when established evaluate *Expression*.

⟦"⊩"$_{Theory}$ aGoal:*Pattern*⟧:*Expression*
Set aGoal for *Theory* and return a list of assertions that satisfy the goal.

⟦**"When"** "⊩"$_{Theory}$ aGoal:*Pattern* "→" *Expression*⟧
When there is a goal that matches aGoal for *Theory*, evaluate *Expression*.

Illustration of backward chaining:

⊢$_t$ Human[Socrates]▮

**When** ⊩$_t$ Mortal[*x*] → (⊩$_t$ Human[$$*x*] → ⊢$_t$ Mortal[*x*])▮

⊩$_t$ Mortal[Socrates]▮

will result in asserting Mortal[Socrates] for theory t.

**SubArguments**

This section explains how subarguments[i] can be implemented in natural deduction.

**When** ⊩$_s$ (*psi* ⊢$_t$ *phi*) →
  **Let** t' ← extension.[t],
  **Do** ⊢$_{t'}$ *psi*,
  ⊩$_{t'}$ *phi* → ⊢$_s$ (*psi* ⊢$_t$ *phi*)▮

Note that the following hold for t' because it is an extension of t:

- **when** ⊢$_t$ *theta* → ⊢$_{t'}$ *theta*▮
- **when** ⊩$_{t'}$ *theta* → ⊩$_t$ *theta*▮

---

[i] See appendix on Inconsistency Robust Natural Deduction.

## Aggregation using Grounded-Complete Predicates

Logic Programs in ActorScript are a further development of Planner. For example, suppose there is a grounded-complete predicate[99] Street[*aName*, *anAddress*, *anotherAddress*, *aDistance*] that is true exactly when the street with *aName*, *anAddress* and *anotherAddress* has *aDistance*.

**When** ⊩ Street[*aName*, *anAddress*, *anAddress*, *aDistance*]→
                    // when a goal is set for a distance between *anAddress* and itself
  ⊢ *aDistance*=0▮                // assert that the distance from an address to itself is 0

The following goal-directed Logic Program works forward from *start* to find the distance to *finish*:

**When** ⊩ Distance[*start*, *finish*, *aDistance*]→
  ⊢ *aDistance*=Minimum.[{*nextDistance* + *remainingDistance*
                | ⊩ Street[*aName*, *start*, *next*, *nextDistance*],
                  Distance[*next*, *finish*, *remainingDistance*]}[100]]▮
    // the distance from *start* to *finish* is the minimum of the set of the sums of the
      // distance for the next address after *start* and
        // the distance from that address to *finish*

The following goal-directed Logic Program works backward from *finish* to find the distance from *start*:

**When** ⊩ Distance[*start*, *finish*, *aDistance*]→
  ⊢ *aDistance*=
      Minimum.[{*remainingDistance* + *previousDistance*
            | ⊩ Street[*aName*, *previous*, *finish*, *previousDistance*],
              Distance[*start*, *previous*, *remainingDistance*]}[101]]▮
    // the distance from *start* to *finish* is the minimum of the set of the sums of the
      // distance for the previous address before *finish* and
        // the distance from *start* to that address

Note that all of the above Logic Programs work together concurrently.

The following procedure computes the shortest path from one location to another:

ShortestPath.[start, finish] ≡
  start �
    finish ⁘ [start],     // the shortest path from start to itself is [start]
    **else** ⁘
     [start, ∨ ⊩ Distance[start,
                 *next*,
                 Minimum.[{*aDistance*
                     | ⊩ Street[_, start, *between*≠start, _]
                        Distance[*between*,
                             finish,
                             *aDistance*]}]] →
          [*next*, ∨ ShortestPath.[*next*, finish]▮

## Appendix 4. ActorScript Symbols with Readings. IDE ASCII, and Unicode code points

| Symbol | IDE ASCII[i] | Read as | Category | Matching Delimiters | Unicode (hex) |
|---|---|---|---|---|---|
| ▮ | ;; | end | top level terminator | | 32DA |
| : | : | of exact type | infix | | |
| :: | :: | of extended type | infix | | |
| ⍠ | [:] | cast this actor to | prefix | | 2360 |
| ∆ | /_\ | discriminate | infix | | 2206 |
| ↓ | \/ | resolve | prefix | | 2139 |
| ⊡ | [.] | qualified by | infix | | 22A1 |
| ▪ | . | is sent | infix | | |
| ▪▪ | .. | delegate to this Actor | prefix | | |
| ⦷ | (\|\|) | concurrently [102] | prefix | | 29B7 |
| ↦ | \|-> | message type returns[103] | infix | | 21A6 |
| → | --> | message received[104] | | ¶ | 2192 |
| ← | <-- | be[105] | infix | | 2190 |
| � | <?> | has cases | separator | ⍰ | FFFD |
| ⍰ | [?] | end cases | terminator | � and catch� | 2370 |
| ¶ | \p | another message handler | separator for handlers | → | 00B6 |
| § | \s | end handlers | terminator | **implements** | 00A7 |
| ⦂ | (:) | case | separator for case | | 2982 |
| ⬣ | ; | before | separator | Let bindings, Do preparations, Enqueue, | 00C4 |
| ≡ | === [106] | defined as | infix | | 2261 |
| ≜ | =/\= | to be | infix | | 225C |
| ≔ | := | is assigned | infix | | 2254 |
| $$ | $$ | matches value of[107] | prefix | | |
| = | = | same as? | infix | | |
| ⌸ | [=] | keyword or field | infix | | 2338 |

[i] These are only examples. They can be redefined using keyboard macros according to personal preference.

| | | | | | |
|---|---|---|---|---|---|
| :⌸ | :[=] | assignable field | infix | | |
| ◁ | <\| | begin type parameters | left delimiter | ▷ (Unicode hex: 0077) | 0076 |
| ⩛ | \\|/ | spread[108] | prefix | | 2A5B |
| { | { | begin set | left delimiter | } | |
| [ | [ | begin list | left delimiter | ] | |
| ⦃ | {\| | begin multi-set | left delimiter | ⦄ | 2983 |
| ⟦ | [\| | array reference | left delimiter | ⟧ (Unicode hex: 27E7) | 27E6 |
| ( | ( | begin grouping | left delimiter | ) | |
| ⦅ | (\| | begin syntax | left delimiter | ⦆ (Unicode hex: 2986) | 2985 |
| ⊝ | (-) | nothing[109] | expression | | 229D |
| ↞ | <<– | one-way send | infix | | 219E |
| ↠ | –>> | one-way receive | infix | ¶ | 21A0 |
| ⊔ | \|_\| | join | infix | | 2294 |
| ⊑ | [<=] | constrained by | infix | | 2291 |
| ⊒ | [>=] | extends | infix | | 2292 |
| ⇒ | ==> | logical implication | infix | | 21E8 |
| ⇔ | <=> | logical equivalence | infix | | 21D4 |
| ∧ | /\ | logical conjunction | infix | | 00D9 |
| ∨ | \/ | logical disjunction | infix | | 00DA |
| ¬ | -\| | logical negation | prefix | | 00D8 |
| ⊢ | \|- | assert | prefix and infix | | 22A2 |
| ⊩ | \|\|- | goal | prefix and infix | | 22A9 |
| // | // | begin 1-line comment | prefix | EndOfLine | |
| /* | /* | begin comment | prefix | */ | |

### Appendix 5. Grammar Precedence

**In the diagram below, if there is no precedence relationship, then parentheses *must* be used.**

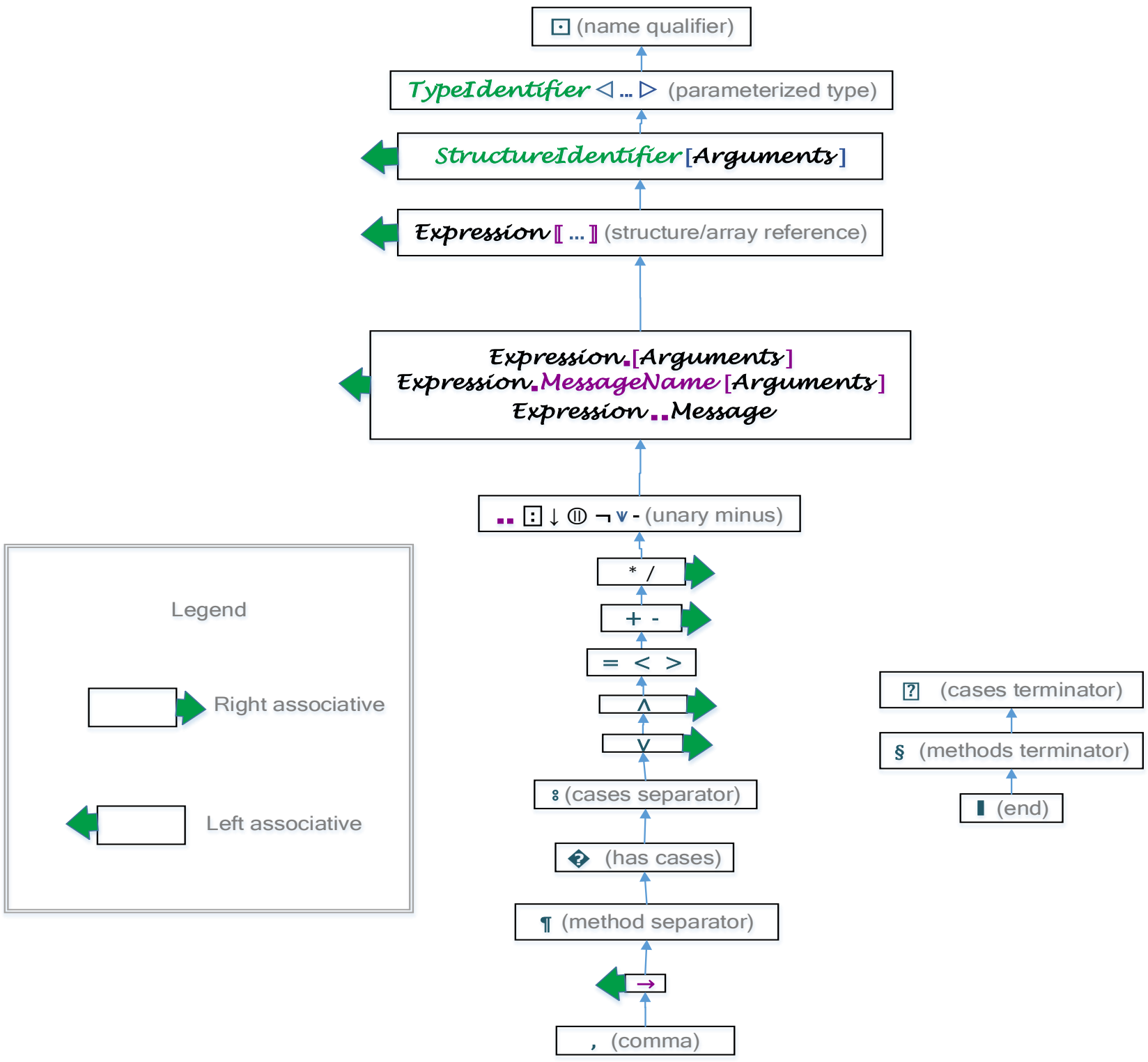

For example, parentheses *must* be used in the following examples:

- (t[p]).m[x]
- (x � p1 ⁝ y1 ⍰) � p2 ⁝ y2 ⍰

# Index

---

**End Notes**

[1] Quotation by the author from late 1960s.

[2] to use a reserved word as an identifier it could prefixed, *e.g.,* _actor

[3] The delimiters ⦅ and ⦆ are used to delimit program syntax with the character “ and the character ” to delimit tokens. For example, ⦅3 “+” 4⦆ is an expression that can be evaluated to 7. A special font is used for syntactic categories. For example *Numerical* is the syntactic category of numerical expressions. The character ⊑ can be used to make categorizations.

For example, ⦅x:*Numerical* “+” y:*Numerical*⦆:*Numerical*▮ where *Numerical*⊑*Expression*▮ with x and y that are both of category *Numerical*.

Also,

⦅*Numerical* “-” *Numerical*⦆:*Numerical*▮
⦅“-” *Numerical*⦆:*Numerical*▮
⦅*Numerical* “*” *Numerical*⦆:*Numerical*▮
⦅*Numerical* “/” *Numerical*⦆:*Numerical*▮
⦅“**Remainder**” *Numerical* “/” *Numerical*⦆:remainder:*Numerical*▮
⦅“**QuotientRemainder**” *Numerical*
“/” *Numerical*⦆:*QuotientRemainderExpression*▮
*QuotientRemainderExpression*⊑*Expression*▮
⦅“**True**” ⊔ “**False**” ⦆:*Expression*◁**Boolean**▷▮
⦅*Expression*◁**Boolean**▷ “∧” *Expression*◁**Boolean**▷⦆
:*Expression* ◁**Boolean**▷▮
⦅*Expression* “∨” *Expression*⦆:*Expression*◁**Boolean**▷▮
⦅ “¬” *Expression*◁**Boolean**▷⦆:*Expression*◁**Boolean**▷▮
⦅ “**Throw**” *Expression*⦆:*Expression*▮

[4] See explanation of syntactic categories above. A word must begin with an alphabetic character and may be followed by one or more numbers and alphabetic characters.

*Identifier* ⊑*Word*⊑*Expression*▮
// an *Identifier* is a *Word*, which is a subcategory of *Expression*
⦅⦅*Expression* ⊔ *Definition* ⊔ *Judgment*⦆⦆ “▮”⦆:*Top*▮

[5] *Type*⊑*Expression*◁**Type**▷▮
⦅ aType:*Type* “↦” anotherType:*Type*⦆:*Type*▮
⦅ “[” *Types* “]” ⦆:*Type*▮
⦅ ⊔ *MoreTypes*⦆:*Types*▮
⦅*Type* ⊔ ⦅*Type* “,”*MoreTypes*⦆⦆:*MoreTypes*▮

[6] ⦅*Identifier*◁**aType**▷ ⦅ ⊔ ⦅“:” *Type*◁**aType**▷⦆⦆
“≡” *Expression*◁**aType**▷⦆:*Definition*◁**Identifier**▷▮

[7] An overloaded procedure is one that takes different actions depending on the types of its arguments.

[8] Note the Symbols box provided by an Integrated Development Environment (IDE) above to make it easier to construct a program by selecting symbols from a context sensitive picker. Also an IDE can automatically provide syntax completion alternatives Analogous to **ctrl+space** in Eclipse, *etc.*.

[9] ⦅“[” *ArgumentTypes* “]”
“↦” returnType:*TypeExpression*⦆:*ProcedureSignature*▮
*ProcedureSignature*⊑*Expression*▮
// signature for a procedure with *ArgumentTypes* and returnType
⦅ ⊔ *MoreArgumentTypes*⦆:*ArgumentTypes*▮
⦅*TypeExpression*
⊔ ⦅*TypeExpression* “,” *MoreArgumentTypes*⦆⦆
:*MoreArgumentTypes*▮
⦅“**Interface**” *Identifier*◁**Type**▷
“{” *MoreProcedureSignatures* “}”⦆:*ProcedureInterface*▮
⦅*ProcedureSignature*
⦅ ⊔ *MoreProcedureSignatures*⦆⦆:*MoreProcedureSignatures*▮

[10] ⦅“[” *ArgumentDeclarations* “]” ⦅ ⊔ ⦅ “:” *Type*◁**aType**▷ ⦆⦆
⦅ ⊔ ⦅“**sponsor**” *Identifier*◁**Sponsor**▷⦆⦆
“→” *Expression*◁**aType**▷⦆:*Procedure*▮
*Procedure*⊑*Expression*▮
// Procedure with *ArgumentDeclarations* that returns
// *Expression* of type returnType.
⦅ ⊔ *MoreDeclarations*⦆:*ArgumentDeclarations*▮
⦅*SimpleDeclaration*⦅ ⊔ ⦅“,” *MoreKeywordDeclarations*⦆⦆
⊔ ⦅*SimpleDeclaration* “,” *MoreDeclarations*⦆⦆
:*MoreDeclarations*▮
// Comma is used to separate declarations.
⦅⦅*Identifier*
⊔ ⦅*Identifier* “:” *Expression*◁**Type**▷ ⦆⦆
⦅ ⊔ “**default**” *Expression*⦆⦆:*SimpleDeclaration*▮
⦅*KeywordArgumentDeclaration*
⊔ ⦅*KeywordDeclaration* “,”*MoreKeywordDeclarations*⦆⦆
:*MoreKeywordDeclarations*▮
⦅*Keyword* “▤”*SimpleDeclaration*⦆⦆:*KeywordDeclaration*▮
*Keyword*⊑*Word*▮

[11] The symbol **.** is fancy typography for an ordinary period when it is used to denote message sending.

[12] ⦅Recipient:*Expression* “**.**” “[” *Arguments* “]” ⦆:*ProcedureSend*▮
*ProcedureSend*⊑*Expression*▮
// Recipient is sent a message with *Arguments*
⦅ ⊔ *MoreArguments*⦆:*Arguments*▮

⦅⦅*Expression* ⦅ ⊔ ⦅"," *MoreKeywordArguments*⦆⦆⦆
⊔ ⦅*Expression* "," *MoreArguments*⦆⦆:*MoreArguments*▮
⦅*KeywordArgument*
⊔ ⦅*KeywordArgument*
"," *MoreKeywordArguments*⦆⦆:*MoreKeywordArguments*▮
⦅*Keyword* "⊟" *Expression*⦆:*KeywordArgument*▮
⦅*Identifier*◁Procedure▷ "▪" "["*ArgumentDeclarations* "]"
⦅ ⊔ ⦅":" returntype:*Type*◁aType▷⦆⦆
"≡" *Expression*◁aType▷ "▮"⦆:*Definition*◁Procedure▷▮

13 ⍰ takes care of the infamous "dangling else" problem [Abrahams 1966].

14 ⦅test:*Expression* "�"
*ExpressionCases*◁aType▷ "⍰"⦆:*Expression*◁aType▷▮
⦅*ExpressionCase*◁aType▷
⊔ *MoreExpressionCases*◁aType▷⦆:*ExpressionCases*◁aType▷▮
⦅*ExpressionCase*◁aType▷ ⊔
⦅*ExpressionCase*◁aType▷ "," *MoreExpressionCases*◁aType▷⦆
⊔ *ExpressionElseCases*◁aType▷⦆ :*MoreExpressionCases*◁aType▷▮
⦅ ⊔ *ExpressionElseCase*◁aType▷
⊔ ⦅*ExpressionElseCase*◁aType▷
"," *MoreExpressionElseCases*◁aType▷⦆⦆
:*ExpressionElseCases*◁aType▷▮
⦅*ExpressionElseCase*◁aType▷
⊔ ⦅*ExpressionElseCase*◁aType▷
"," *MoreExpressionElseCases*◁aType⦆⦆
:*MoreExpressionElseCases* ◁aType▷▮
⦅ ⦅"**else**" "⸪" *Expression*◁aType▷⦆
⊔ ⦅"**else**" *Pattern*"⸪" *Expression*◁aType▷⦆⦆
:*ExpressionElseCase*◁aType▷▮
// The else case is executed only if the patterns before
// the else case do not match the value of test.
⦅*Pattern*"⸪" *Expression*◁aType ⦆:*ExpressionCase*◁aType▷▮

15 ⦅"**Let**" "{" *MoreConcurrentLetBindings* "}"⦅"," ⊔ "⬣"⦆
result:*Expression*◁aType▷⦆:*Expression*◁aType▷▮
// Bindings are independent of each other
⦅"**Let**" "[" *MoreDependentLetBindings* "]"⦅"," ⊔ "⬣"⦆
result:*Expression*◁aType▷⦆:*Expression*◁aType▷▮
// Each binding is dependent on previous ones

⦅"**Let**" *LetBinding* ⦅"," ⊔ "⬣"⦆ result:*Expression*◁aType▷⦆
:*Expression*◁aType▷▮
⦅*LetBinding* ⊔ ⦅*LetBinding* "," *MoreConcurrentLetBindings* ⦆⦆
:*MoreConcurrentLetBindings*▮
⦅*LetBinding*

⊔ ⦅LetBinding ⦅“,” ⊔ “⬣”⦆ MoreDependentLetBindings ⦆⦆
:MoreDependentLetBindings ▮

// Each binding before a “⬣” is completed before its successors

⦅Pattern “←” Expression⦆:LetBinding ▮

[16] Dijkstra[1968] famously blamed the use of the goto as a cause and symptom of poorly structure programs. However, assignments are the source of much more serious problems.

[17] The example could be written as follows using fewer symbols:

```
Actor Account[startingBalance:Currency]
   myBalance ≔ startingBalance
   getBalance[ ]→ myBalance
   deposit[anAmount]→
        Void afterward myBalance ≔ myBalance + anAmount
   withdraw[anAmount]→
      (amount > myBalance) �
         True ⦂ Throw OverdrawnException[ ]
         False ⦂
           Void afterward myBalance ≔ myBalance – anAmount▮
```

[18] ⦅“Actor” Constructor ActorBody⦆:Expression ▮

// The above expression creates an Actor with

// declarations for variables and message handlers

⦅⦅ ⊔ ⦅ “extends” Constructor⦆⦆

⦅ ⊔ “management” Expression◁[aType]↦Manager▷⦆

NamedDeclaration

MessageHandlers

InterfaceImplementations⦆:ActorBody ▮

⦅ActorQueues NamesDeclarations⦆:NamedDeclaration ▮

⦅ ⊔ ⦅MoreNameDeclarations⦆⦆:NamesDeclarations ▮

⦅NameDeclaration

⊔ ⦅NameDeclaration

“,” MoreNamesDeclarations⦆⦆:MoreNameDeclarations ▮

⦅Identifier

⦅ ⊔ ⦅“:”Type◁aType▷⦆⦆

“←” Expression◁aType▷⦆:IdentifierDeclaration ▮

IdentifierDeclaration⊑NameDeclaration ▮

⦅Variable⦅ ⊔ ⦅“:”Type◁aType▷⦆⦆

“≔” Expression◁aType▷ InstanceVariableAQualifications⦆
:VariableDeclaration ▮

VariableDeclaration⊑NameDeclaration ▮

Variable⊑Word ▮

InstanceIVariableQualifications⊑ InstanceQualifications ▮

⦅ ⊔ InstanceVariableQualification

⊔ ⦅ *InstanceVariableQualification*
*InstanceIVariableQualifications* ⦆
:*InstanceIVariableQualifications* ▮
"**nonpersistent**"⊑*InstanceVariableQualification* ▮
// A nonpersistent variable must be of type **Nullable**◁***aType***▷,
// and can be nulled out before a message is received
⦅ ⊔ ⦅ "**queue**" *QueueName* ⦆ ⊔ ⦅ "**queues**" *QueueNames* ⦆⦆
:*ActorQueues* ▮
⦅*QueueName* ⊔ ⦅*QueueName* "**,**" *QueueNames* ⦆⦆:*QueueNames* ▮
*QueueName* ⊑*Word* ▮
*QueueName* ⊑*Expression* ◁**Queue**▷▮
⦅"**Void**"⦆:*Expression* ▮

[19] ⦅recipient:*Expression*
"**.**" *MessageName* "**[**" *Arguments* "**]**"⦆:*NamedMessageSend* ▮
*NamedMessageSend* ⊑*Expression* ▮
// Recipient is sent message *MessageName* with *Arguments*
*MessageName* ⊑*Word* ▮
⦅"**Interface**" *Identifier*◁**Type**▷
⊔ ⦅"**extends**" *Identifier*◁**anotherType**▷
"**{**" *MoreMethodSignatures* "**}**"⦆
⦅ "**{**" *MoreMethodSignatures* "**}**"⦆:*ActorInterface* ▮
*ActorInterface* ⊑*Definition* ▮
⦅*MethodSignature* ⦅ ⊔ *MoreMethodSignatures* ⦆⦆
:*MoreMethodSignatures* ▮
⦅*MessageName* "**[**" *ArgumentTypes* "**]**" "**↦**"
returnType:*TypeExpression* ⦆:*MethodSignature* ▮
*MethodSignature* ⊑*Expression* ▮

[20] Continuations in ActorScript are related to continuations introduced in [Reynolds 1972] in that they represent a continuation of a computation. The difference is that a continuation of Reynolds is a procedure that takes as an argument the result of the preceding computation. Consequently, a continuation of Reynolds is closer to a customer in the Actor Model of computation.

[21] ⦅*InterfaceImplementation* ⦅ ⊔ *InterfaceImplementations* ⦆⦆
:*InterfaceImplementations* ▮
⦅⦅ ⊔ "**partially**"⦆
⦅"**implements**" ⊔ "**reimplements**"⦆ *Type* ◁**aType**▷ "**using**"
⦅*Methods* "**§**"⦆⊔ *UniversalMethod* ⦆
:*InterfaceImplementation* ◁**aType**▷▮
⦅*MessagePattern*
⦅ ⊔ ⦅"**:**" *Type* ⦆⦆
⦅ ⊔ ⦅"**sponsor**" *Identifier*◁**Sponsor**▷⦆⦆

"→" *Continuation*◁aType▷⟧:*UniversalMethod*◁aType▷▮
⟦ ⊔ *MoreMethods*⟧:*Methods*▮
⟦*Method* ⊔ ⟦*Method* "¶" *MoreMethods*⟧⟧:*MoreMethods*▮
// The method separator is ¶.
⟦*MessageName* "[" *ArgumentDeclarations* "]"
⟦ ⊔ ⟦ ":" returnType:*Type*◁aType▷⟧
⟦ ⊔ ⟦"**sponsor**" *Identifier*◁Sponsor▷⟧⟧
"→" *Continuation*◁aType▷⟧:*Method*▮
// For a message with *MessageName* with arguments,
// the response is *Continuation*
⟦*Expression*◁aType▷
"**afterward**" "{" *Afterward* "}"⟧:*Continuation*◁aType▷▮
// Return *Expression* and afterward perform
// *MoreVariableAssignments*
⟦*VariableAssignment*
⊔ ⟦*VariableAssignment*
"," *MoreVariableAssignments*⟧⟧
:*MoreVariableAssignments*▮
// Variable assignments are separated using ","
⟦*Variable* ":=" *Expression*◁aType▷⟧:*VariableAssignment*◁aType▷▮
22 ⟦"Ⓣ" anExpression:*Expression*◁aType▷
⟦ ⊔ ⟦"**sponsor**" *Expression*◁Sponsor▷▮⟧⟧:*Expression*◁aType▷▮
// Execute anExpression concurrently and respond with the outcome.
// In every case, anExpression must complete before execution leaves the lexical scope in which it appears.
23 ⟦"**Do**" "{" *MoreConcurrentAntecedents* "}" ⟦"," ⊔ "⬣"⟧
*Continuation*◁aType▷")" ⟧:*Continuation*◁aType▷▮
⟦"**Do**" "[" *MoreSequentialAntecedents* "]" ⟦ ⊔ "⬣"⟧
*Continuation*◁aType▷")" ⟧:*Continuation*◁aType▷▮
⟦"**Do**" *Antecedent* ⟦"," ⊔ "⬣"⟧
*Continuation*◁aType▷")" ⟧:*Continuation*◁aType▷▮
⟦"**Let**" "{" *MoreConcurentLetBindings* "}" ⟦"," ⊔ "⬣"⟧
result:*Continuation*◁aType▷ ⟧:*Continuation*◁aType▷▮
⟦"**Let**" "[" *MoreDependentLetBindings* "]" ⟦"," ⊔ "⬣"⟧
result:*Continuation*◁aType▷ ⟧:*Continuation*◁aType▷▮
⟦*Antecedent* ⊔ ⟦ *Antecedent* "," *MoreConcurrentAntecedents*⟧⟧
:*MoreConcurrentAntecedents*▮
⟦*Antecedent* ⊔ ⟦ *Antecedent* ";" *MoreSequentialAntecedents*⟧⟧
:*MoreSequentialAntecedents*▮
*Expression*⊑*Antecedent*▮
*StructureAssignment*⊑*Antecedent*▮
*ArrayAssignment*⊑*Antecedent*▮

[24] Swiss cheese was called "serializers" in the literature.

[25] Sussman and Steele 1975] introduced the name "continuation passing style" for explicit use of continuations [Reynolds 1972] in programs. However, in the "string bean style" used here, continuations are not made explicit while programs are required to be linear between holes in the cheese.

[26] without leaving the cheese provided that **aQueue** is nonempty

[27] without leaving the cheese provided that **aQueue** is nonempty

[28] ⦅"**Enqueue**" passThruQ:*Expression* "⬣"
⦅ ⊔ ⦅ "**backout**" backout:*Expression* ⦆⦆
*Continuation* ◁**aType**▷⦆:*Continuation*◁**aType**▷▮
Enqueue this activity in passThruQ. When the cheese is re-entered, perform *Continuation*. If this activity is removed from the queue, execute the backout expression.

[29] **Precondition** requires that a prerequisite be met else an exception is thrown. See description later in this article.

[30] because it is an internal error condition that should never occur

[31] without leaving the cheese provided that **aQueue** is nonempty

[32] without leaving the cheese provided that **aQueue** is nonempty

[33] ⦅*Expression* ◁**aType**▷ "**where**" *Definition* ⦆:*Expression*◁**aType**▷▮

[34] **Interface** Counter {stop[ ]↦Integer,
go[ ]↦Void}▮

[35] ⦅"**Enqueue**" *QueueExpression* ⦅ ⊔ "**after**" "{" *Preparation* "}"⦆ "⬣"
*Continuation* ◁**aType**▷⦆⦆⦆:*Continuation* ◁**aType**▷▮

1. Perform *Preparation*
2. Enqueue activity in *QueueExpression*
3. Leave the cheese
4. When the cheese is re-entered perform *Continuation*.

⦅"▪▪" *Message* ◁**aType**▷⦆:*Expression* ◁**aType**▷▮
Delegate message to this Actor.

Cases can be continuations:

⦅test:*Expression* "�"
*ContinuationCases* ◁**aType**▷ "⍰"⦆:*Continuation* ◁**aType**▷▮
⦅*ContinuationCase* ◁**aType**▷
⊔ ⦅*ContinuationCase* ◁**aType**▷
"," *MoreContinuationCases* ◁**aType**▷⦆⦆
*ContinuationElseCases*⦆:*ContinuationCases*▮
⦅*ContinuationCase* ◁**aType**▷
⊔ ⦅*ContinuationCase* ◁**aType**▷
"," *MoreContinuationCases* ◁**aType**▷⦆⦆
:*MoreContinuationCases* ◁**aType**▷▮
⦅*Pattern*"⦂" *Continuation* ◁**aType**▷⦆:*ContinuationCase* ◁**aType**▷▮
⦅ ⊔

*MoreContinuationElseCases* ◁aType▷⟫
:*ContinuationElseCases* ◁aType▷▮
⟪*ContinuationElseCase* ◁aType▷
⊔ ⟪*ContinuationElseCase* ◁aType▷
"," *MoreContinuationElseCases* ◁aType▷⟫⟫
:*MoreContinuationElseCases* ◁aType▷▮
⟪⟪"else" "⦂" *Continuation* ◁aType▷⟫
⊔ ⟪"else" *Pattern* "⦂"
*Continuation*◁aType▷⟫⟫:*ContinuationElseCase* ◁aType▷▮

[36] Popularized in [Dijkstra 1965]. Note that secondBarber only shaves if a customer comes in when firstBarber is shaving.

[37] **Interface** BarberShop visit[Client]↦ Void▮

[38] because it is an internal error condition that should never occur

[39] after shaving permit next customer

[40] leave cheese while shaving after recording that first barber is shaving

[41] after shaving permit next customer

[42] ReadersWriterConstraintMonitor defined below monitors a resource and throws an exception if it detects that ReadersWriter constraint is violated, *e.g.*, for a resource r using the above scheduler, ReadingPriority[ReadersWriterConstraintMonitor[r]].

```
Actor ReadersWriterConstraintMonitor[theResource:ReadersWriter]
  writing ≔ False,
  numberReading:(Integer thatIs ≥0) ≔ 0,
  implements ReadersWriter using
   read[query]→
     Precondition ¬writing;
       Hole  theResource.read[query]
         after numberReading++

         afterward numberReading--¶
   write[update]→
     Preconditions numberReading = 0, ¬writing;
        Hole  theResource.write[update]
         after writing≔True

         afterward writing ≔ False §▮
```

43 A downside of this policy is that readers may not get the most recent information.

44 A downside of this policy is that writing and reading may be delayed because of lack of concurrency among readers.

45 Precondition that is present for inconsistency robustness.

46 ++ is postfix increment

47 -- is postfix decrement

48 Precondition that is present for inconsistency robustness.

49 The following are allowed in the cheese for a response to message affecting the next message:

⦅*Expression* ◁**aType**▷
⦅ ⊔ ⦅ "**permit**" aQueue:*Expression*⦆⦆
⦅ ⊔ ⦅"**afterward**" "**{**" *Afterward* "**}**"⦆⦆⦆:*Continuation* ◁**aType**▷▮

If there are activities in aQueue, then the one of them gets the cheese next and also perform *Afterward*; then leave the cheese and return the value of *Expression*.

The following can be used temporarily leave the cheese:

⦅"**Hole**" *Expression*◁**aType**▷⦆:*Continuation* ◁**aType**▷▮

1. Leave the cheese
2. The response is the result of evaluating *Expression*

⦅"**Hole**" *Expression*◁**aType**▷
⦅ ⊔ ⦅"**after**" *Preparation*⦆⦆⦆:*Continuation* ◁**aType**▷▮

1. Carry out *Preparation*
2. Leave the cheese
3. The result is the result of evaluating *Expression*

⦅"**Hole**" *Expression*◁**aType**▷
⦅ ⊔ ⦅"**after**" *Preparation*⦆⦆
⦅ ⊔ ⦅ "**afterward**" *Afterward*⦆:*Continuation* ◁**aType**▷▮

1. Carry out *Preparation*
2. Leave the cheese
3. Evaluate *Expression*
4. When a response is received, reacquire the cheese, carry out *Afterward* and the result is the result of evaluating *Expression*

If *Expression* throws an exception, continue using the exception *ContinuationCases*.

⦅"**Hole**" *Expression*◁**anotherType**▷
⦅ ⊔ ⦅"**after**" *Preparation*⦆⦆
⦅ ⊔ ⦅ "**returned�**" *ContinuationCases* ◁**aType**▷ "⍰"⦆⦆
⦅ ⊔ ⦅ "**threw�**" *ContinuationCases* ◁**aType**▷ "⍰"⦆⦆
:*Continuation* ◁**aType**▷▮

1. Carry out *Preparation*
2. Leave the cheese
3. Evaluate *Expression*
4. When a response is received, reacquire the cheese
   - If *Expression* returns, continue using the returned Actor with *ContinuationCases*
   - If *Expression* throws an exception, continue using the exception *ContinuationCases*.

50 ⦅*Identifier*◁**Type**▷<br>
  "◁" *ParametersDeclarations* "▷"<br>
    "≡" *Expression*⦆:*ParameterizedDefinition* ▮<br>
*ParameterizedDefinition*⊑*Definition* ▮<br>
  Parameterize definition with *ParametersDeclarations* ▮<br>
⦅ ⊔ *MoreParameterDeclarations* ⦆:*ParametersDeclarations* ▮<br>
⦅*ParameterDeclaration*<br>
  ⊔ ⦅*ParameterDeclaration*<br>
    "," *MoreParameterDeclarations*⦆⦆<br>
      :*MoreParameterDeclarations* ▮<br>
⦅*Identifier*◁**Type**▷ ⦅ ⊔ *Qualifier* ⦆⦆:*ParameterDeclaration* ▮<br>
⦅ ⊔ ⦅"**extends**" *Identifier*◁**Type**▷ ⦆⦆:*TypeQualifier* ▮<br>
⦅*Identifier*◁**Type**▷ "◁" *Parameters* "▷" ⦆:*TypeExpression* ▮<br>
⦅*Identifier*◁**Type**▷<br>
  ⊔ ⦅ ⊔ ⦅*Identifier*◁**Type**▷ "," *Parameters* ⦆⦆:*Parameters* ▮<br>
51 ⦅*Identifier*◁**aType**▷ "[" *Arguments* "]"⦆:*Expression* ◁**aType**▷ ▮<br>
⦅*Identifier*◁**aType**▷ "[" *Patterns* "]"⦆:*Pattern* ◁**aType**▷ ▮

52 ⦅"**Structure**" *Identifier*◁**Type**▷ "[" *FieldDeclarations* "]"<br>
    *StructureImplementation* ⦆:*Definition* ▮<br>
  // Structure definition with *StructureImplementation*<br>
⦅ ⊔ *MoreFieldDeclarations* ⦆:*FieldDeclarations* ▮<br>
⦅⦅*SimpleFieldDeclaration*<br>
      ⦅ ⊔ ⦅"," *MoreNamedFieldDeclarations* ⦆⦆⦆<br>
  ⊔ ⦅*SimpleFieldDeclaration*<br>
    "," *MoreFieldDeclarations*⦆⦆:*MoreFieldDeclarations* ▮<br>
  // Comma is used to separate declarations.<br>
⦅⦅*Identifier*<br>
  ⊔ ⦅*Identifier* ":" *TypeExpression*⦆⦆<br>
    ⦅ ⊔ "**default**" *Expression*⦆⦆:*SimpleFieldDeclaration* ▮<br>
⦅*NamedFieldDeclaration*<br>
  ⊔ ⦅*NamedFieldDeclaration*<br>
    "," *MoreNamedFieldDeclarations*⦆⦆<br>
      :*MoreNamedFieldDeclarations* ▮<br>
⦅*FieldName*<br>
  ⦅"□" ⊔ ":□"⦆ *SimpleFieldDeclaration*⦆⦆

:*NamedFieldDeclaration*▮

*FieldName*⊑*QualifiedName*▮

// ":⌸" is used for assignable fields.

⦅⦅ ⊔ *Identifier*⦆ *ActorBody*⦆:*StructureImplementation*▮

⦅*Expression* "⟦" *FieldName* "⟧" ⦆:*FieldSelector*▮

// *FieldName* of *Expression* which *must* be a structure

*FieldSelector*⊑*Expression*▮

⦅*StructureName* "[" *FieldExpressions* "]" ⦆:*StructureExpression*▮

*StructureExpression*⊑*Expression*▮

⦅ ⊔ *MoreFieldExpressions*⦆:*FieldExpressions*▮

⦅⦅*SimpleFieldExpression*⦅ ⊔ ⦅"," *MoreNamedFieldExpressions*⦆⦆⦆

⊔ ⦅*SimpleFieldExpression*

"," *MoreFieldExpressions*⦆⦆:*MoreFieldExpressions*▮

⦅*NamedFieldExpression*

⊔ ⦅ *NamedFieldExpression*

"," *MoreNamedFieldExpressions*⦆⦆

:*MoreNamedFieldExpressions*▮

⦅*FieldName*

⦅"⌸" ⊔ ":⌸"⦆ *SimpleFieldExprression*⦆⦆

:*NamedFieldExpression*▮

⦅*StructureName* "[" *FieldPatterns* "]" ⦆:*StructurePattern*▮

*StructurePattern*⊑*Pattern*▮

⦅ ⊔ *MoreFieldPatterns*⦆:*FieldPatterns* ▮

⦅⦅*SimpleFieldPattern*⦅ ⊔ ⦅"," *MoreNamedFieldPatterns*⦆⦆⦆

⊔ ⦅ *SimpleFieldPattern* "," *MoreFieldPatterns*⦆⦆

:*MoreFieldPatterns*▮

⦅*NamedFieldPattern*

⊔ ⦅ *NamedFieldPattern*

"," *MoreNamedFieldPatterns*⦆⦆

:*MoreNamedFieldPatterns*▮

⦅*FieldName* ⦅"⌸" ⊔ ":⌸"⦆ *SimpleFieldExprression*⦆⦆

:*NamedFieldPattern*▮

53 ⦅"**Try**" anExpression:*Expression*◁**aType**▷

"**catch**�" *ExpressionCases*◁**aType**▷ "⍰"⦆:*Expression*◁**aType**▷▮

- If anExpression throws an exception that matches the pattern of a case, then the value of *TryExpression* is the value computed by *ExpressionCases*
- If anExpression doesn't throw an exception, then then the value of *TryExpression* is the value computed by anExpression.

⦅"**Try**" anExpression:*Expression*◁**aType**

"**catch**�" *ContinuationCases*◁**aType**▷"⍰"⦆

:*Continuation*◁**aType**▷▮

- If anExpression throws an exception that matches the pattern of a case, then the response of *TryContinuation* is the response computed by the expression of the case.
- If aContinuation doesn't throw an exception, then then the response of *TryExpression* is the response computed by anExpression.

⦅"**Try**" anExpression:*Expression*◁**aType**▷
"**cleanup**" cleanup:*Expression*◁**aType**▷⦆:*Expression*◁**aType**▷▮

- If anExpression throws an exception, then the value of *TryExpression* is the value computed by cleanup.
- If anExpression doesn't throw an exception, then then the value of *TryExpression* is the value computed by anExpression.

54 ⦅"**Precondition**" *Expressions* ";" *Expression*⦆:*Expression*▮
// Each of *Expressions* must evaluate to **True** or an exception is thrown
⦅"**Precondition**" *Expressions* ";" *Continuation*⦆:*Continuation*▮
// Each of *Expressions* must evaluate to **True** or an exception is thrown
⦅value:*Expression*◁**aType**▷
"**postcondition**" pre:*Expression*◁[**aType**]↦**Boolean**▷⦆
:*Expression*◁**aType**▷▮
// The expression pre must evaluate to **True** when sent value
// or an exception is thrown

55 º is a reserved postfix operator for degrees of angle

56 i.e., i*i=-1 where i is the imaginary number **Cartesian**[0, 1]

57 ⦅"[" *ComponentExpressioons* "]"⦆:*Expression*◁**List**▷▮
// An ordered list with elements *Expressions*
⦅ ⊔ *MoreComponentExpressioons*⦆:*ComponentExpressioons* ▮
⦅⦅⦅ ⊔ "∀"⦆ *Expression*⦆
⊔ ⦅⦅ ⊔ "∀"⦆ *Expression*
"," *MoreComponentExpressioons*⦆⦆
:*MoreComponentExpressioons*▮
⦅"[" *TypeExpressions* "]"⦆:*TypeExpression*▮
⦅ ⊔ *MoreTypeExpressions*⦆:*TypeExpressions*▮
⦅*TypeExpression* ⊔ ⦅*TypeExpression*
"," *MoreTypeExpressions*⦆⦆
:*MoreTypeExpressions*▮

58 ⦅"_"⦆:*UnderscorePattern*▮
*UnderscorePattern*⊑*Pattern*▮
*Identifier*⊑*Pattern*▮
⦅*Pattern*◁**aType**▷":" *Type*◁**aType**▷⦆:*Pattern*◁**aType**▷▮
⦅*Pattern* "**suchThat**" *Expression*⦆:*SuchThat*▮
*SuchThat*⊑*Pattern*▮
⦅*Pattern* "**thatIs**" *Expression*⦆:*ThatIs*▮
*ThatIs*⊑*Pattern*▮
⦅"$$" *Expression*◁**Type**▷⦆:*Pattern*◁**Type**▷▮

⦅"[" *ComponentPatterns* "]"⦆:*Pattern*◁List▷▮
// A pattern that matches a list whose elements match
// *ComponentPatterns*
⦅ ⊔ *MoreComponentPatterns* ⦆:*ComponentPatterns*▮
⦅*Pattern*
⊔ ⦅ "⋎"*Pattern* ⦆
⊔ ⦅*Pattern* "," *MoreComponentPatterns* ⦆⦆
:*MoreComponentPatterns*▮

59 Equivalent to the following:

AlternateElements.[aList:List◁aType▷]:List◁aType▷ ≡
aList �
List◁aType▷[ ] ⦂ List◁aType▷[ ],
List◁aType▷[anElement] ⦂ List◁aType▷[anElement],
List◁aType▷[firstElement, secondElement] ⦂
List◁aType▷[firstElement],
else
List◁aType▷[firstElement,
secondElement,
⋎remainingElements] ⦂
List◁aType▷[firstElement,
⋎AlternateElements.[remainingElements]] ?▮

60 ⦅"{" *ComponentExpressioons* "}"⦆:*Expression*◁Set▷▮
A set of Actors without duplicates
⦅"{" *ComponentPatterns* "}"⦆:*Pattern*◁Set▷▮

61 ⦅"⦃" *ComponentExpressioons* "⦄"⦆:*Expression*◁Multiset▷▮
A multiset of the Actors with possible duplicates
⦅"⦃" *ComponentPatterns* "⦄"⦆:*Pattern*◁Multiset▷▮

62 Optimization of this program is facilitated because:

- The records are determinate because their type is Set◁ContactRecord[62]▷
- All of the operators return determinate results

- The operators are annotated as **determinate**

[63] ⦅“Map” “{” *ComponentExpressioons* “}” ⦆:*Expression*◁Map▷▮

[64] It is possible to define a procedure that will produce a “bottomless” future.
For example, f.[ ]:Future◁aType▷ ≡ **Future** f.[ ]▮

[65] the examples using ⓟ can be slightly more efficient *as written*

[66] ⦅**Postpone** *Expression*◁aType▷▮⦆:*Expression*◁Future◁aType▷▷▮
postpone execution of an expression until the value is needed.

[67] ⦅“**Future**” aValue:*Expression*◁aType▷
⦅ ⊔ ⦅“**sponsor**” *Expression*◁Sponsor▷⦆⦆⦆
:*Expression*◁Future◁aType▷▷▮
A future for aValue.

⦅“↓” *Expression*Future◁aType▷▷⦆:*Expression*◁aType▷▷▮
Resolve a future

[68] ⦅LoopName:*Identifier* “.” “[” *Initializers* “]”
⦅ ⊔ ⦅ “:” ReturnType:aType ⦆⦆
“≙” *Expression*◁aType▷ ⦆:*Expression*◁aType▷▮

⦅ ⊔ *MoreInitializers*⦆:*Initializers*▮

⦅*Initializer* ⊔ ⦅*Initializer* “,” *MoreInitializers*⦆⦆
:*MoreInitializers*▮

⦅*Identifier* ⦅ ⊔ ⦅“:” *TypeExpression*⦆⦆ “←” *Expression*⦆:*Initializer*▮

[69] ⦅“**Discrimination**” *Identifier*◁Type▷
“{”*MoreTypeDescriminations*“}“ ⦆:*Expression*◁Type▷▮

⦅*Identifier*◁Type▷
⊔ ⦅*Identifier*◁Type▷
“,”*MoreTypeDescriminations*⦆⦆
:*MoreTypeDescriminations*▮

⦅*Expression*◁aDiscriminationType▷ “Δ” *Type*◁aType▷⦆
:*Expression*◁aType▷▮
// Discriminate to have the type *Type*◁aType▷ if possible.
// Otherwise, an exception is thrown.

[70] The implementation below requires careful optimization.

[71] ⦅“String” “[” *ComponentExpressioons* “]”⦆:*Expression*◁String▷▮

⦅“String” “[” *ComponentPatterns* “]”⦆:*Pattern*◁String▷▮

[72] ⦅recipient:*Expression*◁recipientType▷
“.” message:*MessageExpression*◁recipientType▷⦆:*Expression*▮
Send recipient the message

[73] ⦅ “⦅” *MoreGrammers* “⦆” ⦆:*Grammar*▮

⦅ “⦅”*Grammar*“⊔”*Grammar*“⦆” ⦆:*Grammar*▮

⦅*ReservedWord* ⦅ ⊔ *StartsWithIdentifier* ⦆⦆:*StartsWithReserved*▮

*StartsWithReserved* ⊑*MoreGrammers*▮

⦅*Identifier* ⦅ ⊔ *StartsWithReserved* ⦆⦆⦆:*StartsWithIdentifier*▮

*StartsWithIdentifier* ⊑*MoreGrammers*▮

⦅"\"" *Word* "\""⦆ :*ReservedWord* ▮
// The use of \ escapes the next character in a string so
// that "\"" has just one character that is ".
⦅*Grammar* ":" *GrammarIdentifier* "▮"⦆:*Judgment* ▮
⦅*Identifier*◁**Grammar**▷ "⊑" *Identifier*◁**Grammar**▷ "▮"⦆:*Judgment* ▮

74 The implementation below can be highly inefficient.

75 ⦅"**Atomic**" aLocation:*Expression*
"**compare**" comparison:*Expression*
"**update**" update:*Expression*
"**then**" compareIdentical:*Continuation*◁**aType**▷
"**else**" compareNotIdenticial:*Continuation*◁**aType**▷⦆
:*Continuation* ◁**aType**▷▮
Atomically compare the contents of location with the value of comparison. If identical, update the contents of aLocation with the value of update and execute compareIdentical.

76 ⦅*Identifier* "⊡" *Qualifier* ⦆:*QualifiedName* ▮
*QualifiedName* ⊑*Expression* ▮
*Identifier* ⊑*QualifiedName* ▮
⦅*Identifier* ⊔ ⦅*Identifier* "⊡" *Qualifier* ⦆⦆:*Qualifier* ▮

77 ⦅"**Enumeration**" *Identifier*◁**Type**▷
"{" *MoreEnumerationNames* "}" ⦆:*Definition* ▮
⦅*EnumerationName*
⊔ ⦅*EnumerationName*
"," *MoreEnumerationNames*⦆⦆
:*MoreEnumerationNames* ▮
*EnumerationName* ⊑*Word* ▮

78 Declarations provide version number, encoding, schemas, *etc.*

79 If a customer is sent more than one response (i.e., **return** or **throw** message) then it will throw an exception to the sender of the response.

80 ⦅recipient:*Expression*
"↞" *MessageName* "[" *Arguments* "]"⦆:*Expression* ◁**Void**▷▮
// recipient is sent one-way
// message *MessageName* with *Arguments*
// Note that *Expression* ◁⊖▷ *cannot* be used to produce a value

81 ⦅*MessageName* "[" *ArgumentDeclarations* "]"
⦅ ⊔ ⦅"**sponsor**" *Identifier*◁**Sponsor**▷ ⦆⦆▷⦆⦆
"↠" *Continuation* ◁⊖▷⦆:*Method* ▮
// one-way method implementation
// with *ArgumentDeclarations* that has a
// one-way continuation that returns nothing
⦅"⊖" ⦅ ⊔ ⦅"**permit**" aQueue:*Expression*⦆⦆
⦅ ⊔ ⦅" **afterward**" *Assignments*⦆⦆⦆:*Continuation* ◁"⊖"▷▮

82 Hoare[1962]. The implementation below is adapted from Wikipedia.

[83] Move Actor at pivotIndex to end

[84] **Interface** Account◁aCurrency⊑Currency▷
{getBalance[ ] ↦ aCurrency,
deposit[aCurrency] ↦ Void,
withdraw[aCurrency] ↦ Void}▮

[85] cf. [Crahen 2002, Amborn 2004, Miller, et. al. 2011]

[86] **Interface** AccountMonitor◁aCurrency▷
{getRevoker[ ] ↦ AccountRevoker,
getAccount[ ] ↦ Account
withdrawFee[aCurrency] ↦ Void}▮

[87] **Interface** AccountRevoker◁aCurrency▷
{revokeDeposit[ ] ↦ Void,
revokeWithdraw[ ] ↦ Void}▮

[88] Consider a dialect of Lisp which has a simple conditional expression of the following form:

⦅"(" "**if**" test:*Expression* then:*Expression* else:*Expression* ")" ⦆

which returns the value of then if test evaluates to **True** and otherwise returns the value of else.

The definition of Eval in terms of itself might include something like the following [McCarthy, Abrahams, Edwards, Hart, and Levin 1962]:

```
(Eval expression environment) ≡
                                   // Eval of expression using environment defined to be
 (if (Numberp expression)                        // if expression is a number then
    expression                                        // return expression else
    (if ((Equal (First expression) (Quote if))
                                                // if First of expression is "if" then
       (if (Eval (First (Rest expression) environment)
                                  // if Eval of First of Rest of expression is True then
          (Eval (First (Rest (Rest expression)) environment)
                               // return Eval of First of Rest of Rest of expression else
          (Eval (First (Rest (Rest (Rest expression)) environment))
                             // return Eval of First of Rest of Rest of Rest of expression
    ...))
```

The above definition of Eval is notable in that the definition makes use of the conditional expressions using **if** expressions in defining how to Eval an **If** expression!

[89] The implementation CheeseQ uses activities to implement its queue where for type Activity the following holds:

**Structure** Activity[previous :⊟ Nullable◁Activity▷,
// if null then head of queue else,
// pointer to backwards list to head
nextHint :⊟ Nullable◁Activity▷]▮
// if non-null then pointer to next
// activity to get cheese after this one

[90] if non-null points to head with current holder of cheese

[91] if non-null, pointer to backwards list ending with head that holds cheese

[92] **Interface** CheeseQ {enter[ ] ↦ Void,
leave[ ] ↦ Void}▮

[93] **enter** message received running myActivity

[94] this cheese queue is not empty because myActivity is at the head of the queue

[95] **Interface** SubCheeseQ head[ ] ↦ Activity▮

[96] **Interface** InternalQ {enqueueAndLeave[ ] ↦ **Void**,
enqueueAndDequeue[InternalQ] ↦ Activity,
dequeue[ ] ↦ Activity,
isEmpty[ ] ↦ Boolean}▮

[97] **Interface** subInternalQ {add[Activity] ↦ **Void**,
remove[ ] ↦ Activity}▮

[98] [Church 1932; McCarthy 1963; Hewitt 1969, 1971, 2010; Milner 1972, Hayes 1973; Kowalski 1973]. Note that this definition of Logic Programs does *not* follow the proposal in [Kowalski 1973, 2011] that Logic Programs be restricted only to clause-syntax programs.

[99] A grounded-complete predicate is one for which all instances in which the predicate holds are explicitly manifest, i.e. instances can be generated using patterns. See [Ross and Sagiv 1992, Eisner and Filardo 2011].

[100] Execution can proceed differently depending on whether this set fits in memory.

[101] Execution can proceed differently depending on whether this set fits in memory.

[102] following expression is executed concurrently

[103] Used in type specifications for interfaces.

[104] Used in methods.

[105] Used to bind identifiers in **Let**.

[106] Three equal signs because two equal signs have a meaning in Java

[107] Used in patterns.

[108] Used in structures.

[109] Used in one-way message passing.